\documentclass[showpacs,floatfix,superscriptaddress]{revtex4}
\usepackage{graphicx}
\usepackage{caption}
\usepackage{subfig}
\usepackage{epsfig}
\usepackage{bm}
\usepackage[utf8]{inputenc}
\usepackage{amssymb}
\usepackage{float}
\usepackage{amsmath}
\usepackage{dcolumn}
\usepackage{latexsym}
\usepackage{subfig}
\usepackage{amsthm}
\usepackage{framed}
\usepackage{cancel}
\usepackage[colorlinks]{hyperref}
\usepackage[usenames,dvipsnames]{color}
\usepackage{multirow}
\usepackage[pdftex]{pict2e}
\usepackage[dvipsnames]{xcolor}
\usepackage{subfig}

\hypersetup{
     breaklinks=true,
    pdfstartview={FitH},    
    colorlinks=true,       
    linkcolor=blue,          
    citecolor=red,        
    filecolor=magenta,      
    urlcolor=blue,           
    anchorcolor=green,      
    linktocpage=true
}

\def\doi{http://doi.org}

\begin{document}

\title{Higher Dimensional Loop Quantum Black hole in de Sitter Spacetime: Quasinormal Modes and Shadow Signatures}

\author{Sara Saghafi}
\email[]{saghafisara1366@gmail.com}
\affiliation{Department of Theoretical Physics, Faculty of Science, University of Mazandaran,\\
P. O. Box 47416-95447, Babolsar, Iran}

\author{Kourosh Nozari}
\email[]{nozari7450@gmail.com, knozari@umz.ac.ir (Corresponding Author)}
\affiliation{Department of Theoretical Physics, Faculty of Science, University of Mazandaran,\\
P. O. Box 47416-95447, Babolsar, Iran}

\author{Ali Mohammadpour}
\email[]{alimohammadpour9898@gmail.com}
\affiliation{Department of Theoretical Physics, Faculty of Science, University of Mazandaran,\\
P. O. Box 47416-95447, Babolsar, Iran}

\begin{abstract}
We investigate the dynamical and optical signatures of a
higher-dimensional loop-quantum-corrected black hole in a de Sitter
background. The physically admissible nonextremal solutions are
restricted to a finite mass interval bounded by the extremal and
Nariai configurations. We find that loop-quantum corrections produce
moderate shifts in the scalar quasinormal spectrum, whereas increasing
the spacetime dimension has a substantially stronger effect, enhancing
both the oscillation frequency and the damping rate. All calculated
modes have negative imaginary parts, supporting dynamical stability
against massless scalar perturbations in the parameter range considered.
Comparison with the corresponding classical backgrounds shows that the
quantum-corrected spectrum remains continuously connected to the
classical photon-sphere branch, with the quantum correction producing
predominantly quantitative modifications. The black-hole shadow is
also sensitive to the loop-quantum parameter, the cosmological
constant, and the spacetime dimension, with extra dimensions generally
reducing its size. Comparison with the Event Horizon Telescope
measurement of M87$^*$ constrains the allowed model parameters.
Together, these results show that quasinormal modes and shadow
observables provide complementary probes of loop-quantum corrections
and higher-dimensional effects in the strong-field regime.

\end{abstract}


\maketitle

 \section{INTRODUCTION}

Spacetime singularities constitute one of the fundamental limitations of
classical general relativity. The singularity theorems of Penrose and
Hawking--Penrose demonstrate that, under appropriate causality, energy,
and trapped-surface conditions, gravitational collapse generically leads
to geodesic incompleteness \cite{Penrose1965,HawkingPenrose1970,hawking2023large}.
These results indicate that classical general relativity cannot provide
a complete description of spacetime in sufficiently strong-gravity
regimes. An additional fundamental issue concerns whether the
singularities formed during gravitational collapse are necessarily
hidden behind event horizons. This question is closely related to the
cosmic censorship conjecture \cite{Wald1997}, according to which
physically reasonable gravitational collapse is expected not to produce
singularities visible to distant observers. The possible occurrence of
naked singularities would therefore have important consequences for the
causal and predictive structure of classical general relativity.\\

A consistent theory of gravitation is expected to address, at a fundamental level, the emergence and resolution of spacetime singularities. This requirement naturally motivates the development of a quantum theory of gravity that unifies the principles of quantum mechanics with those of general relativity. Among the various candidates, loop quantum gravity (LQG) has emerged as a particularly compelling non-perturbative and background-independent framework, within which substantial progress has been achieved (see, e.g., \cite{rovelli1995discreteness,ashtekar1997quantum,han2007fundamental,ashtekar2006quantum,ashtekar2011loop,zhang2022first,zhang2023fermions} and references therein). By implementing loop quantization techniques in symmetry-reduced spacetimes, especially in the context of spherically symmetric black holes, important insights into the quantum structure of black hole interiors have been obtained \cite{chiou2008phenomenological,gambini2008black,haggard2015quantum,christodoulou2016realistic,ashtekar2018quantum,zhang2020loop,zhang2022loop,
lewandowski2023quantum,husain2022fate}. In these approaches, the classical Schwarzschild singularity is generically resolved due to quantum geometric effects, although the precise mechanism of singularity resolution depends on the specific quantization scheme employed. Furthermore, quantum geometry effects in loop quantum cosmology suggest that the classical big bang singularity is replaced by a non-singular bounce, thereby providing a consistent picture of early-universe dynamics \cite{stachowiak2007exact,ashtekar2006quantum2}. These developments have also led to the construction of improved effective black hole models incorporating quantum corrections \cite{modesto2006loop,bojowald2018signature,chiou2008phenomenological2}. In particular, within the quantum Oppenheimer--Snyder collapse model formulated in loop quantum cosmology, a novel class of quantum black hole solutions has recently been obtained \cite{lewandowski2023quantum}, in which the classical singularity is replaced by a transition region bounded by an inner horizon. Based on such quantum-corrected geometries, several studies have investigated observable signatures, including black hole shadows and stability properties \cite{zhang2023black,yang2023shadow}, revealing notable deviations from classical expectations in asymptotically flat spacetimes. On the other hand, mounting observational evidence for the positively accelerated expansion of the late time universe points toward the presence of a positive cosmological constant. This motivates the inclusion of a de Sitter background when analyzing quantum-corrected black holes \cite{shao2024scalar}. Consequently, it becomes essential to examine how the presence of a cosmological constant modifies the response of such black holes to perturbations, as well as their observable characteristics.\\

The exploration of higher-dimensional spacetimes constitutes an active and conceptually rich area of modern theoretical physics. A principal motivation for considering extra dimensions arises from the longstanding goal of unifying gravity with the other fundamental interactions within a single theoretical framework \cite{Horava:1995qa}. While the Standard Model of particle physics successfully describes three of the four known fundamental interactions, it does not incorporate gravitation. Higher-dimensional theories, such as Kaluza--Klein models \cite{kaluza1921unitatsproblem,klein1999quantum,bailin1987kaluza} and M-theory \cite{witten1996five}, offer a natural setting in which gauge fields and matter content can emerge from the geometry of additional spatial dimensions, thereby providing a pathway toward unification. Beyond this, extra-dimensional frameworks have been invoked to address a range of outstanding cosmological problems, including the origin of dark matter and the nature of dark energy \cite{dvali20004d,qiang2005five,zhang2013loop}. Despite the progress achieved in loop quantum gravity, most investigations of loop quantum black hole models have been restricted to four-dimensional spacetimes. In this work, we extend the de Sitter quantum Oppenheimer--Snyder model \cite{shao2024scalar} to higher dimensions in order to construct a non-singular quantum-corrected Schwarzschild black hole solution in a higher-dimensional de Sitter background. This generalization enables us to explore how the interplay between quantum geometry effects and extra dimensions influences the physical properties of black holes.\\

A fundamental aspect of black hole spacetimes is their stability under external and internal perturbations \cite{regge1957stability}. Perturbations of black holes can be broadly classified into external (exogenous) disturbances, arising from surrounding matter or fields, and internal (endogenous) fluctuations associated with the intrinsic properties of the black hole itself, such as its mass, angular momentum, or charge. The dynamical response of a perturbed black hole typically proceeds through three distinct stages: an initial transient phase, a quasinormal mode (ringdown) phase, and a late-time tail. Among these, the quasinormal mode (QNM) phase is of particular importance, as it is characterized by damped oscillations with complex frequencies determined solely by the background spacetime geometry. These frequencies are independent of the details of the initial perturbation, rendering QNMs as intrinsic signatures of the black hole. Consequently, they provide a powerful probe of the underlying spacetime structure and play a central role in gravitational-wave phenomenology \cite{leaver1986spectral,berti2009quasinormal,de2026confronting}. QNMs provide a powerful diagnostic for probing the dimensional structure of spacetime. In particular, in scenarios with extra spatial dimensions—such as those arising in braneworld models—the properties of black holes may deviate significantly from their four-dimensional counterparts, and higher-dimensional black holes could, in principle, emerge in sufficiently high-energy gravitational regimes \cite{abdalla2007perturbations,konoplya2007stability,zhidenko2008evolution,abdalla2007quasinormal}. Over the past decades, the study of QNMs in higher-dimensional black hole spacetimes has attracted considerable attention for several reasons. These include gaining insight into the structure of higher-dimensional extension of general relativity \cite{bizon2005critical,bizon2005vacuum,panotopoulos2020quasinormal,chabab2016behavior}, exploring the phenomenological consequences of extra-dimensional scenarios such as braneworld models \cite{kanti2004black} and investigating connections with black hole thermodynamics and quantum gravity frameworks, including loop quantum gravity \cite{hod1998bohr,kunstatter2003d}. In light of the increasing interest in gravitational-wave astronomy and the role of QNMs as observable signatures of black holes, it is particularly relevant to compute the QNM spectrum of higher-dimensional de Sitter black holes incorporating loop quantum corrections. Such an analysis can provide valuable information about how extra dimensions and quantum gravitational effects modify the dynamical response of black holes.\\

In addition to QNMs, considerable attention has recently been devoted to the study of black hole shadows \cite{raza2024influence,atamurotov2023quantum,nozari2025investigating,nozari2025accretion,zhong2021qed,nozari2023asymptotically,aktar2025shadows}. The shadow of a black hole is determined by the behavior of null geodesics in the vicinity of the event horizon and therefore encodes direct information about the underlying spacetime geometry. As such, it provides a powerful probe of the gravitational field in the strong-field regime. A major breakthrough in this direction was achieved by the Event Horizon Telescope (EHT), which produced the first image of the supermassive black hole at the center of the galaxy M87, thereby confirming one of the most striking predictions of general relativity \cite{akiyamaetal2019eventhorizontelescope,event2019first3,akiyama87event,akiyama2019first,event2019first,event2019first2}. These observations have opened a new observational window for testing gravitational theories through black hole shadow measurements. An important feature of EHT observations is that they directly probe the spacetime geometry without relying on a specific gravitational theory, making them a robust tool for distinguishing between general relativity and alternative theories of gravity \cite{psaltis2019testing}. Consequently, black hole shadow observables have been extensively investigated in the literature \cite{jafarzade2025optical,nozari2026rotating,vagnozzi2023horizon,battista2026shadow,capozziello2025null}, particularly in the context of strong-gravity phenomena near the event horizon. A large body of work has focused on understanding how additional degrees of freedom arising in extended theories of gravity—beyond the standard black hole parameters—modify the shadow characteristics \cite{perlick2022calculating}. In this context, the presence of extra spatial dimensions provides an intriguing avenue for exploring deviations from the four-dimensional description of gravity. In principle, observational data from EHT can be employed to probe the existence of extra dimensions and constrain their properties. Previous studies indicate that extra dimensions tend to reduce the size of the black hole shadow across a variety of gravitational models \cite{amarilla2012shadow,eiroa2018shadow,papnoi2014shadow,singh2018shadow,amir2018shadows,belhaj2020deflection}, although the precise impact depends on the underlying theoretical framework. Motivated by these developments, in the present work we aim to investigate whether the mutual effects of extra dimensions and loop quantum corrections can be detected through black hole shadow observations. In particular, we analyze how the combined influence of extra dimensions, a cosmological constant, and loop quantum corrections modifies the shadow radius and compare our results with EHT observations. This approach allows us to assess the extent to which higher-dimensional signatures may be observable and to place constraints on the underlying parameters of the model.\\

The remainder of this paper is organized as follows. In Sec.~II, we construct the quantum-corrected black hole solution in a de Sitter background and extend it to higher-dimensional spacetimes, highlighting the role of loop quantum corrections and extra dimensions in the metric structure. In Sec.~III, we investigate the dynamical properties of the spacetime by analyzing the QNMs of massless scalar perturbations using complementary numerical approaches, including time-domain evolution, the matrix method, and the WKB approximation. In Sec.~IV, we study the optical appearance of the black hole by examining null geodesics and deriving the corresponding shadow observables, followed by a comparison with Event Horizon Telescope observations of M87$^{\ast}$ to constrain the model parameters. Finally, Sec.~V summarizes our main results and discusses their physical implications.

\section{Quantum-Corrected Black Hole in de Sitter Spacetime in the Presence of Extra Dimensions}

We begin by considering the quantum-corrected black hole model introduced in Ref.~\cite{lewandowski2023quantum}, in which the spacetime is divided into two distinct regions. The interior region is described by a homogeneous distribution of pressureless matter (cosmic dust), while the exterior region corresponds to a vacuum spacetime. This construction provides a convenient framework for incorporating quantum gravitational effects through an effective cosmological dynamics in the interior, while maintaining a static geometry outside the matter distribution.

In the interior region, comoving coordinates of the form
$(\tau,\tilde r,\theta,\tilde\phi)$ are introduced, with
\begin{equation}
0\leq \tilde r\leq \tilde r_0,
\end{equation}
where $\tilde r_0$ denotes the fixed comoving radial coordinate of the
boundary of the homogeneous dust ball in the
Ashtekar--Pawlowski--Singh (APS) spacetime. Here, $a(\tau)$ denotes the
scale factor governing the dynamical evolution of the interior geometry.
The corresponding physical (areal) radius of the dust surface is
$a(\tau)\tilde r_0$. During the collapse and subsequent bounce, this
physical radius satisfies $a(\tau)\tilde r_0\in[r_b,\infty)$, where
$r_b$ is the minimum radius attained at the bounce.

The metric inside the dust sphere is then given by a Friedmann--Lemaître--Robertson--Walker (FLRW)-type line element,
\begin{equation}
\label{eq:28_en}
ds_{\text{in}}^{2}
=
- d\tau^{2}
+ a^{2}(\tau)
\left(
d\tilde{r}^{2}
+ \tilde{r}^{2} d\Omega^{2}
\right),
\end{equation}

Before introducing the effective dynamics of the collapsing interior, it is useful to briefly summarize the ingredients of higher-dimensional loop quantum gravity that are relevant to the present construction. We emphasize that the quantum-gravity input employed here does not arise from a direct quantization of the exterior black-hole geometry. Instead, it originates from higher-dimensional loop quantum cosmology (LQC), which provides the homogeneous and isotropic symmetry-reduced sector of the corresponding loop quantum gravity framework. The resulting effective cosmological dynamics is subsequently used in the quantum Oppenheimer--Snyder construction to determine the exterior black-hole geometry through the junction conditions described in Appendix~A.

In a $D$-dimensional spacetime, with $D-1$ spatial dimensions, the connection formulation of higher-dimensional gravity can be expressed in terms of an $SO(D)$ connection $A_{a}^{IJ}$ and its canonically conjugate momentum $\pi^{a}_{IJ}$. Their fundamental Poisson bracket takes the form
\begin{equation}
\left\{
A_{aIJ}(x),\pi^{bKL}(y)
\right\}
=
2\kappa\gamma\,
\delta_{a}^{b}\,
\delta^{K}_{[I}\delta^{L}_{J]}\,
\delta^{(D-1)}(x-y),
\label{eq:LQG_PB}
\end{equation}
where $\kappa=8\pi G$, $a,b$ denote spatial indices, $I,J,K,L$ are internal indices, and $\gamma$ is the Barbero--Immirzi parameter.

Under the assumptions of spatial homogeneity and isotropy, the gravitational phase space reduces to a pair of canonical variables $(c,p)$ related to the scale factor through
\begin{equation}
p=a^{D-2},
\qquad
c=\gamma\dot{a},
\qquad
\{c,p\}
=
\frac{\kappa\gamma}{D-1}.
\label{eq:LQC_cp}
\end{equation}
Thus, the infinitely many gravitational degrees of freedom of the full theory are reduced to the variables governing the isotropic cosmological background relevant to the interior of the collapsing dust ball.

A central feature of loop quantization is that the connection itself is not promoted directly to a well-defined quantum operator. Instead, the gravitational connection enters through its holonomies along finite edges. At the effective level, this results in the characteristic polymerization of the connection variable,
\begin{equation}
c
\longrightarrow
\frac{\sin(\bar{\mu}c)}{\bar{\mu}},
\label{eq:polymerization}
\end{equation}
where the parameter $\bar{\mu}$ is determined by the minimum nonvanishing eigenvalue of the higher-dimensional area operator.

In the improved-dynamics prescription, the corresponding area-gap scale is characterized by
\begin{equation}
\Delta
=
8\sqrt{D-1}\,\pi\gamma\,(l_p)^{D-1},
\qquad
l_p
=
\sqrt{(D-2)G\hbar},
\label{eq:LQC_area_gap}
\end{equation}
following the convention adopted in Refs.~\cite{Zhang2016,shi2024higher}. The discreteness encoded in $\Delta$ introduces a fundamental quantum-geometrical scale and modifies the gravitational dynamics when the energy density approaches the Planck regime.

The effective Hamiltonian dynamics obtained from this holonomy-corrected formulation leads to the higher-dimensional modified Friedmann equation
\begin{equation}
H^2
=
\frac{16\pi G}{(D-1)(D-2)}
\rho
\left(
1-\frac{\rho}{\rho_c}
\right),
\label{eq:LQC_Friedmann_D}
\end{equation}
where the critical density is given by
\begin{equation}
\rho_c
=
\frac{(D-1)(D-2)}
{16\pi G\gamma^2
\Delta^{2/(D-2)}}.
\label{eq:LQC_rhoc_D}
\end{equation}
The quadratic correction in Eq.~(\ref{eq:LQC_Friedmann_D}) is the characteristic effective loop-quantum modification relevant to the present construction. In the low-density regime, $\rho/\rho_c\ll1$, the classical Friedmann dynamics is recovered, whereas the quantum correction becomes important as the density approaches $\rho_c$. In particular, the vanishing of $H$ at $\rho=\rho_c$ corresponds to a quantum bounce, replacing the singular endpoint of the classical collapse by a nonsingular transition.

For the four-dimensional dust interior considered initially below, the inclusion of a positive cosmological constant gives the effective evolution equation
\begin{equation}
H^{2}
=
\left(\frac{\dot{a}}{a}\right)^{2}
=
\frac{8\pi G}{3}
\rho_{\rm matter}
\left(
1-\frac{\rho_{\rm matter}}{\rho_c}
\right)
+
\frac{\Lambda}{3},
\label{eq:LQC_Friedmann_4D}
\end{equation}
with
\begin{equation}
\rho_c
=
\frac{3}{32\pi G\gamma^2}.
\label{eq:LQC_rhoc_4D}
\end{equation}

The presence of the negative quadratic correction term implies the
existence of a non-singular bounce, replacing the classical singularity
with a finite minimum radius. This effective LQC equation provides the
quantum-corrected interior dynamics entering the quantum
Oppenheimer--Snyder (qOS) construction. The exterior black-hole metric
is then obtained by matching this interior solution to a static and
spherically symmetric exterior geometry through the Darmois--Israel
junction conditions. The details of this matching procedure, together
with its extension to $D$ dimensions, are presented explicitly in
Appendix A.

In contrast to the original construction of Ref.~\cite{lewandowski2023quantum}, we explicitly include a positive cosmological constant $\Lambda$, thereby embedding the model within a de Sitter background. This modification is well motivated by cosmological observations and allows us to investigate black hole solutions in an accelerated expanding universe. The assumption of isotropy in the interior region is maintained for simplicity, although more general anisotropic configurations may lead to richer phenomenology and will be explored in future work.

\bigskip

The exterior region is assumed to be static and spherically symmetric, described by coordinates
\[
(t,r,\theta,\phi),
\]
with the metric ansatz
\begin{equation}
\label{eq:31_en}
ds_{\text{out}}^{2}
=
- f(r)\, dt^{2}
+ \frac{1}{g(r)}\, dr^{2}
+ r^{2} d\Omega^{2}.
\end{equation}

The metric functions $f(r)$ and $g(r)$ are determined by matching the interior and exterior geometries across the boundary of the dust sphere using the Darmois--Israel junction conditions. Following the procedure outlined in Ref.~\cite{lewandowski2023quantum}, one obtains
\begin{equation}
\label{eq:32_en}
f(r)=g(r)
=
1-
\left[
\frac{2GM}{r}
+
\frac{\Lambda r^2}{3}
-
\frac{\alpha G^2M^2}{r^4}
\left(
1+\frac{\Lambda r^3}{6GM}
\right)^2
\right],
\end{equation}
where $\alpha$ is a quantum deformation parameter proportional to the Planck area. This correction term encodes the leading-order effects of loop quantum gravity in the exterior geometry. The metric is valid in the region $r \ge r_{b}$, where $r_{b}$ corresponds to the minimum radius reached during the interior bounce.

\bigskip

We now proceed to generalize this construction to higher-dimensional spacetimes. There are two primary motivations for this extension.. First, higher-dimensional gravity arises naturally in various theoretical frameworks, including string theory and braneworld scenarios. Second, the combined presence of quantum corrections, a cosmological constant, and extra spatial dimensions provides a richer setting in which to explore observable signatures such as quasinormal modes and black hole shadows.

Following Ref.~\cite{shi2024higher}, we consider a static, spherically symmetric spacetime in $D$ dimensions described by the line element
\begin{equation}
\label{eq:33_en}
ds^{2}
=
- h(r)\, dt^{2}
+ \frac{dr^{2}}{f(r)}
+ r^{2} d\Omega_{D-2}^{2},
\end{equation}
where $d\Omega_{D-2}^{2}$ denotes the metric on the unit $(D-2)$-sphere.

Within this framework, the metric functions for a quantum-corrected black hole in higher-dimensional de Sitter spacetime take the form derived in Appendix~A,
\begin{equation}
\label{eq:34_en}
h(r)=f(r)
=
1
-
\left(
\frac{2m}{r^{(D-3)}}
-\frac{4\gamma^2 \Delta^{2/(D-2)} m^2}{r^{4(D-3)}}
\right)
\left(
1
+
\frac{2\Lambda r^{3}}{m(D-1)(D-2)}
\right)^{2}
-
\frac{2\Lambda r^{2}}{(D-1)(D-2)},
\end{equation}
where the remaining parameters are defined as
\begin{equation}
m=\frac{8\pi M}{(D-2)\omega},
\qquad
\omega=
\frac{2\pi^{\frac{D-1}{2}}}
{\Gamma\left(\frac{D-1}{2}\right)} .
\label{14}
\end{equation}
This metric incorporates quantum corrections through the
$\gamma$-dependent term, as well as the effects of the cosmological
constant and spacetime dimensionality. In the limit $D=4$, the above
expression reduces to Eq.~(11), providing a consistency check of the
higher-dimensional generalization.

\bigskip

The geometry described by Eqs.~(\ref{eq:33_en})--(\ref{14}) forms the basis for our subsequent analysis. In particular, the structure of null geodesics and the associated photon sphere in this spacetime will determine both the black hole shadow and the high-frequency quasinormal modes. In the following sections, we exploit this connection to investigate the optical and dynamical properties of the quantum-corrected higher dimensional black hole in a unified framework.

\subsection{Horizon structure and physical parameter domain}
\label{subsec:horizon_structure}

Before proceeding to the perturbative and optical analyses, it is
necessary to identify the region of parameter space in which the
higher-dimensional loop-quantum-corrected geometry actually represents
a black hole. The horizons are determined by the positive real roots of
\begin{equation}
f(r)=0.
\label{eq:horizon_condition}
\end{equation}
Depending on the values of the spacetime dimension $D$, the loop
parameter $\gamma$, the cosmological constant $\Lambda$, and the mass
$M$, the metric function may possess three distinct positive roots,
\begin{equation}
r_{-}<r_{h}<r_{c},
\end{equation}
where $r_{-}$ denotes the inner Cauchy horizon, $r_{h}$ the event
horizon, and $r_{c}$ the cosmological horizon.

For fixed $(D,\gamma,\Lambda)$, the existence of these three horizons
places a nontrivial restriction on the black-hole mass. The lower
boundary of the admissible mass interval is determined by an extremal
configuration in which the inner and event horizons coalesce,
$r_{-}=r_{h}$. The corresponding extremal radius $r_{\rm ext}$ and
extremal mass $M_{\rm ext}$ are obtained from the double-root
conditions
\begin{equation}
f(r_{\rm ext};M_{\rm ext})=0,
\qquad
\left.
\frac{\partial f(r;M_{\rm ext})}{\partial r}
\right|_{r=r_{\rm ext}}=0.
\label{eq:extremal_conditions}
\end{equation}
When $M<M_{\rm ext}$, the event horizon no longer exists as a distinct
root and the corresponding configuration lies outside the
three-horizon black-hole sector considered in this work.

At the opposite boundary, increasing the mass causes the event and
cosmological horizons to approach each other. Their coalescence,
$r_{h}=r_{c}$, defines the Nariai configuration. The corresponding
radius $r_{\rm N}$ and mass $M_{\rm N}$ satisfy
\begin{equation}
f(r_{\rm N};M_{\rm N})=0,
\qquad
\left.
\frac{\partial f(r;M_{\rm N})}{\partial r}
\right|_{r=r_{\rm N}}=0.
\label{eq:nariai_conditions}
\end{equation}
Hence, for every fixed set of $(D,\gamma,\Lambda)$, the nonextremal
three-horizon black-hole sector is restricted to
\begin{equation}
M_{\rm ext}<M<M_{\rm N}.
\label{eq:physical_mass_range}
\end{equation}
Equivalently, in terms of the dimensionless mass ratio,
\begin{equation}
\frac{M_{\rm ext}}{M_{\rm N}}
<
\frac{M_{\rm ext}}{M}
<1.
\label{eq:physical_ratio_range}
\end{equation}

It is important to emphasize that $M_{\rm ext}$ and $M_{\rm N}$ are
not universal constants of the model. Rather, they depend on the
specific values of $D$, $\gamma$, and $\Lambda$. Therefore, whenever
the spacetime dimension or the loop parameter is varied, the critical
masses must be determined again from
Eqs.~(\ref{eq:extremal_conditions}) and
(\ref{eq:nariai_conditions}). A fixed numerical value of $M$ cannot
in general be adopted independently of these parameters without first
checking that the corresponding geometry remains within the
three-horizon black-hole sector.

\begin{figure}[htbp]
\centering
\includegraphics[width=0.72\linewidth]{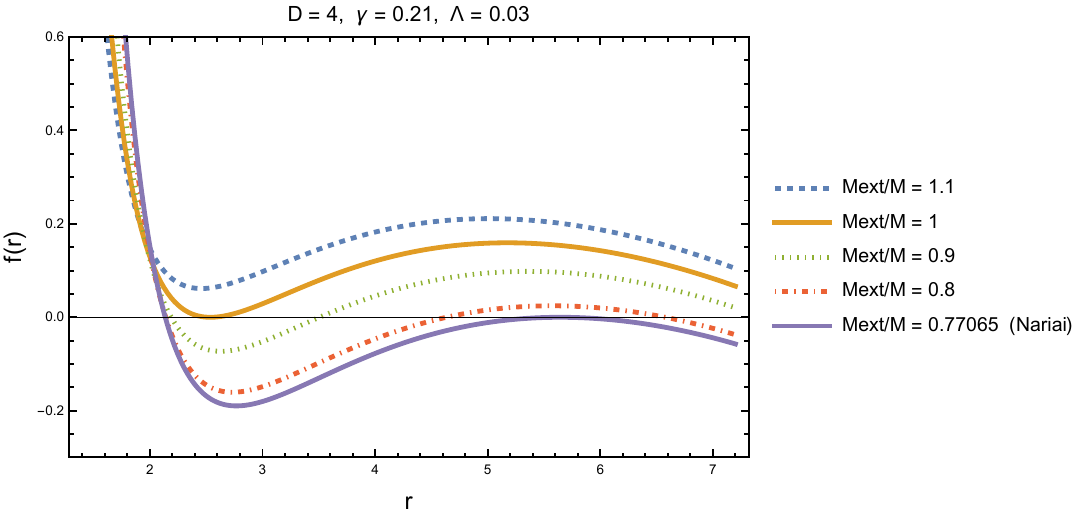}
\caption{\small
Horizon structure of the four-dimensional loop-quantum-corrected
de Sitter black hole for $\gamma=0.21$ and $\Lambda=0.03$.
The metric function $f(r)$ is shown for several values of
$M_{\rm ext}/M$. The extremal configuration
$M_{\rm ext}/M=1$ corresponds to the coalescence of the inner and
event horizons, whereas the Nariai configuration is reached at
$M_{\rm ext}/M\simeq0.77065$, where the event and cosmological
horizons merge. Between these two limiting configurations the metric
possesses three distinct positive roots,
$r_{-}<r_{h}<r_{c}$. The case $M_{\rm ext}/M=1.1$ corresponds to
$M<M_{\rm ext}$ and therefore lies outside the three-horizon
black-hole sector.}
\label{fig:horizonD4}
\end{figure}

\begin{figure}[htbp]
\centering
\includegraphics[width=0.72\linewidth]{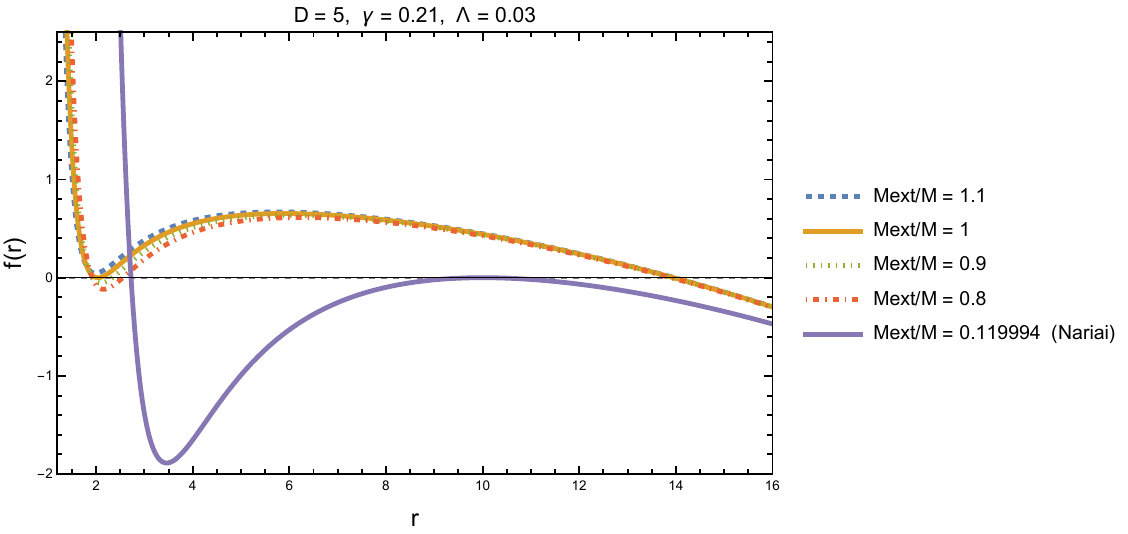}
\caption{\small
Horizon structure of the five-dimensional loop-quantum-corrected
de Sitter black hole for $\gamma=0.21$ and $\Lambda=0.03$.
The extremal configuration occurs at $M_{\rm ext}/M=1$, where the
inner and event horizons coincide. The Nariai limit is reached at
$M_{\rm ext}/M\simeq0.119994$, where the event and cosmological
horizons merge. The intermediate range corresponds to nonextremal
black holes with three distinct horizons,
$r_{-}<r_{h}<r_{c}$.}
\label{fig:horizonD5}
\end{figure}

The different horizon regimes are illustrated in
Figs.~\ref{fig:horizonD4} and \ref{fig:horizonD5}. In both cases,
$M_{\rm ext}/M>1$ corresponds to $M<M_{\rm ext}$ and therefore lies
outside the admissible black-hole domain. At
$M_{\rm ext}/M=1$, the inner and event horizons merge into a
degenerate horizon. For
$M_{\rm ext}/M<1$, the two horizons separate and a nonextremal
three-horizon configuration is obtained until the Nariai limit is
reached, where the event and cosmological horizons coincide.

\begin{figure}[htbp]
\centering
\includegraphics[width=0.78\linewidth]{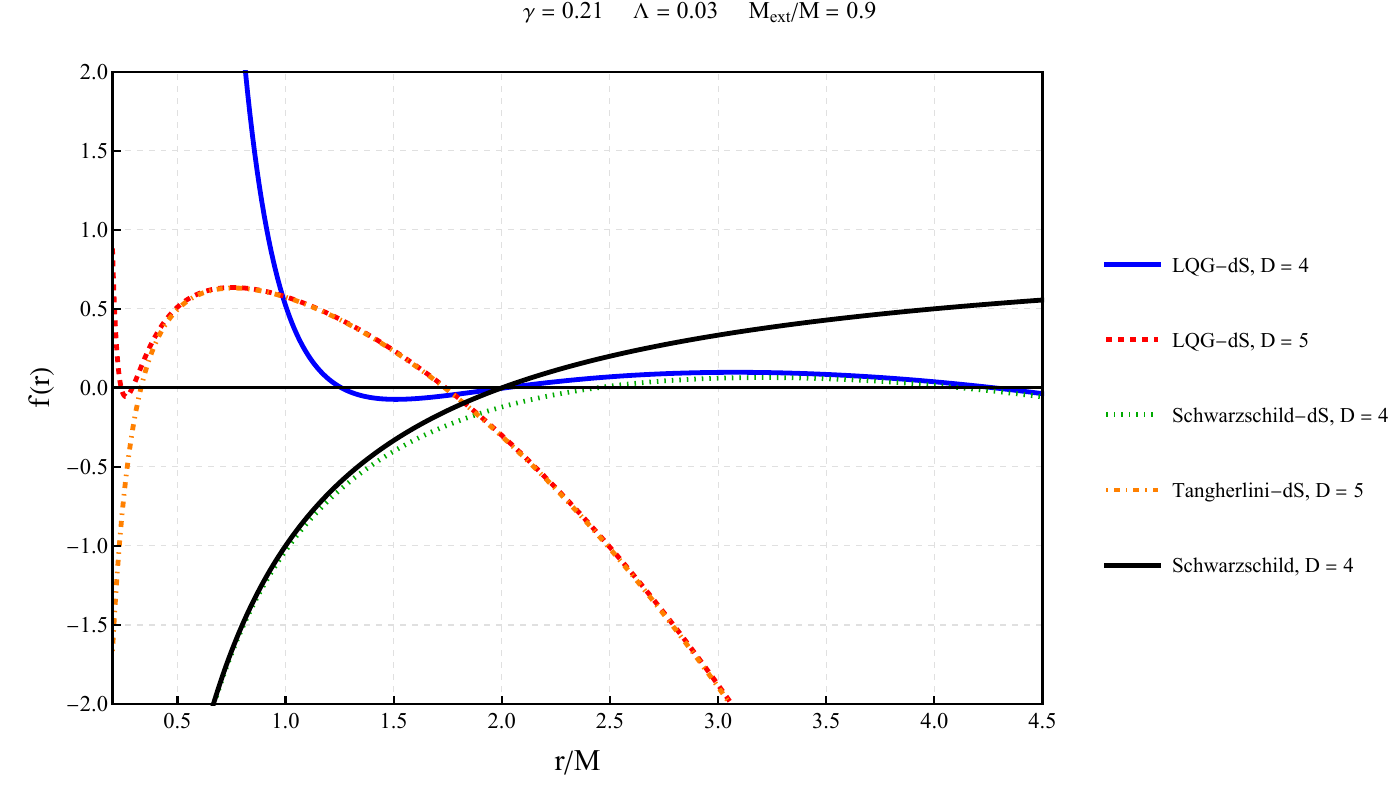}
\caption{\small
Comparison of the metric function $f(r)$ for the
higher dimensional loop-quantum-corrected de Sitter geometries and their classical
counterparts. The quantum-corrected $D=4$ and $D=5$ configurations
are shown for $\gamma=0.21$, $\Lambda=0.03$, and
$M_{\rm ext}/M=0.9$, together with the corresponding
Schwarzschild--de Sitter and Tangherlini--de Sitter backgrounds.
The four-dimensional Schwarzschild solution is also included as the
asymptotically flat reference. The loop-quantum correction modifies
the strong-field profile of the metric and consequently changes the
locations of the horizons relative to the classical solutions.}
\label{fig:metric_comparison}
\end{figure}

To make the departure from the corresponding classical geometries
more explicit, Fig.~\ref{fig:metric_comparison} compares the
loop-quantum-corrected metric functions with the associated classical
backgrounds for a representative nonextremal configuration. For fixed
dimension and cosmological constant, the loop-quantum correction
modifies the strong-field behavior of $f(r)$ and shifts the horizon
locations relative to the Schwarzschild--de Sitter and
Tangherlini--de Sitter cases. The additional Schwarzschild curve
provides the asymptotically flat four-dimensional reference.

For the representative nonextremal choice
$M_{\rm ext}/M=0.9$, the corresponding horizon radii are listed in
Table~\ref{tab:horizon_radii}. This value is chosen sufficiently
away from both the extremal and Nariai boundaries and therefore
provides a convenient configuration for illustrating the subsequent
perturbative analysis.

\begin{table}[htbp]
\centering
\caption{Inner, event, and cosmological horizon radii for the
representative nonextremal configuration
$M_{\rm ext}/M=0.9$, with $\gamma=0.21$ and $\Lambda=0.03$.}
\label{tab:horizon_radii}
\renewcommand{\arraystretch}{1.20}
\begin{tabular}{c|ccc}
\hline
$D$ & $r_{-}$ & $r_{h}$ & $r_{c}$ \\
\hline
4 & 2.190253 & 3.506663 & 7.442389 \\
5 & 1.922184 & 2.334973 & 13.926705 \\
\hline
\end{tabular}
\end{table}

As seen from Table~\ref{tab:horizon_radii}, the expected ordering
\begin{equation}
r_{-}<r_{h}<r_{c}
\end{equation}
is satisfied in both cases. Accordingly, throughout the remainder of
this work we restrict the mass parameter, for every chosen set of
$(D,\gamma,\Lambda)$, to the admissible interval
$M_{\rm ext}<M<M_{\rm N}$. This restriction ensures that all
subsequent calculations are performed on genuine nonextremal
three-horizon black-hole backgrounds. In particular, the same
prescription is adopted in the scalar effective-potential analysis
and in the QNM calculations presented below.
\section{THE QUASINORMAL MODES FOR MASSLESS SCALAR PERTURBATIONS}

Having established the higher-dimensional loop-quantum-corrected de Sitter black hole geometry, we proceed to study its dynamical stability under scalar perturbations. Following the standard treatment employed in Ref.~\cite{shao2024scalar}, we consider a massless neutral scalar field propagating on the background spacetime. The evolution of the scalar perturbation is governed by the covariant Klein--Gordon equation
\begin{equation}
\nabla_{\mu}\nabla^{\mu}\Phi=0.
\end{equation}

Invoking the spherical symmetry of the $D$-dimensional geometry, the scalar field can be separated as
\begin{equation}
\Phi(t,r,\Omega)=\frac{\psi(r)}{r^{\frac{D-2}{2}}}\,Y_{\ell}(\Omega)\,e^{-i\omega t},
\end{equation}
where $Y_{\ell}(\Omega)$ are the hyperspherical harmonics on $S^{D-2}$ and $\ell$ denotes the multipole index. Substituting this decomposition into the Klein--Gordon equation and introducing the tortoise coordinate through
\begin{equation}
dr_{\ast}=\frac{dr}{f(r)},
\end{equation}
the radial equation can be cast into a Schr\"odinger-type wave equation,
\begin{equation}
\left(\frac{d^{2}}{dr_{\ast}^{2}}+\omega^{2}-V_{\mathrm{eff}}(r)\right)\psi(r)=0.
\end{equation}

For a massless scalar in $D$ dimensions, the effective potential takes the form
\begin{equation}\label{eq:Veff}
V_{\mathrm{eff}}(r)=f(r)\left[
\frac{\ell(\ell+D-3)}{r^{2}}
+\frac{(D-2)(D-4)}{4r^{2}}\,f(r)
+\frac{(D-2)}{2r}\,f'(r)
\right],
\end{equation}
so that the loop quantum parameter $\gamma$ influences the perturbation dynamics entirely through the metric function $f(r)$ and its derivatives, modifying the location and curvature of the potential barrier.

\subsection{Effective-potential structure and comparison with
classical black holes}
\label{subsec:potential_comparison}

To make the effect of the loop-quantum correction more transparent,
we now examine the scalar effective potential and compare it with
the corresponding classical black-hole backgrounds. For a
nonextremal de Sitter black hole, the physically relevant static
region for the QNM problem is bounded by the event and
cosmological horizons,
\begin{equation}
r_{h}<r<r_{c}.
\end{equation}
Since $f(r_h)=f(r_c)=0$, Eq.~(\ref{eq:Veff}) implies
\begin{equation}
V_{\rm eff}(r_h)=V_{\rm eff}(r_c)=0.
\end{equation}
The perturbation therefore propagates through an effective
scattering barrier located between these two horizons. Its height,
radial position, and curvature determine the characteristic
oscillation and damping scales entering the quasinormal spectrum.

\begin{figure}[htbp]
\centering
\includegraphics[width=0.78\linewidth]{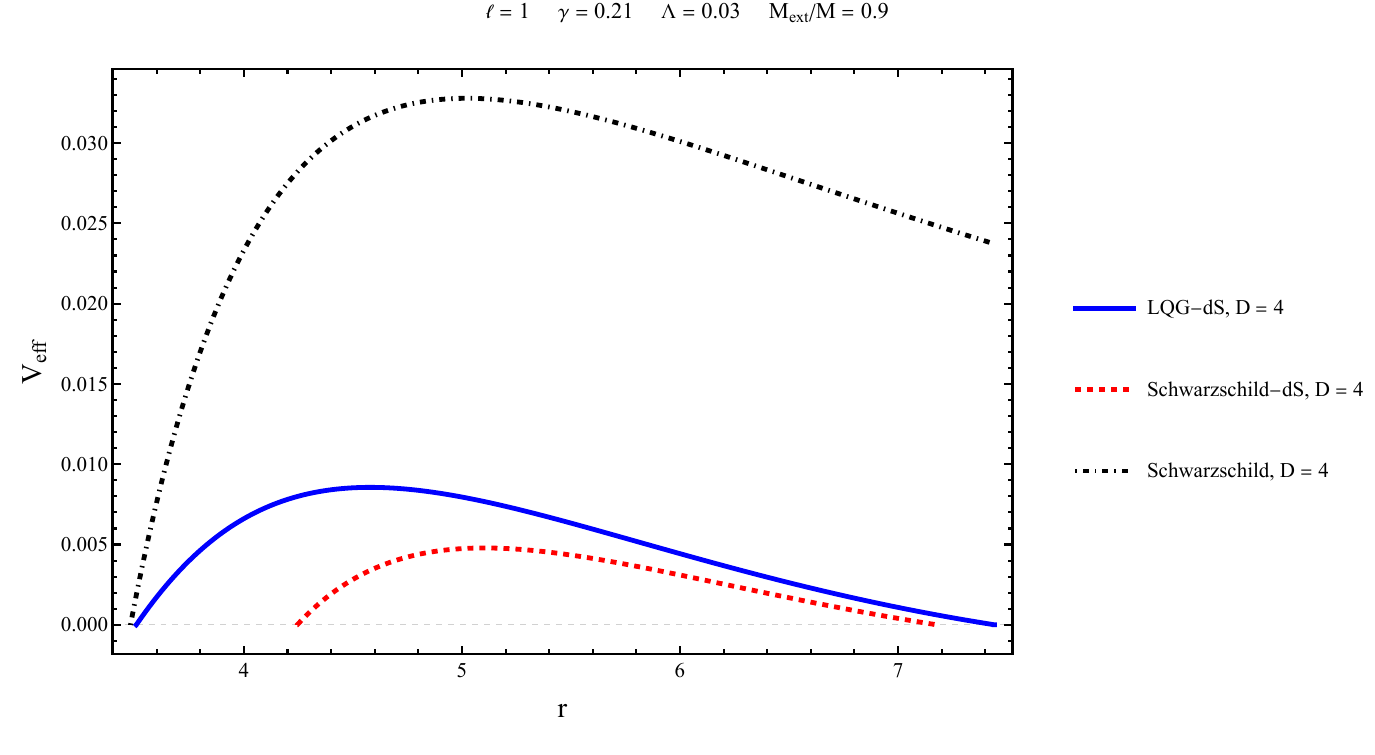}
\caption{\small
Scalar effective potential for the four-dimensional
loop-quantum-corrected de Sitter black hole (LQG--dS), compared with
the Schwarzschild--de Sitter and Schwarzschild backgrounds. We set
$\ell=1$, $\gamma=0.21$, $\Lambda=0.03$, and
$M_{\rm ext}/M=0.9$ for the LQG--dS configuration. For each de Sitter
geometry, the potential is plotted only over its own physical static
region between the corresponding event and cosmological horizons.
Consequently, the LQG--dS and Schwarzschild--de Sitter curves have
different radial endpoints because their horizon locations differ.
The Schwarzschild curve is included as an asymptotically flat
reference.}
\label{fig:potentialD4}
\end{figure}

Figure~\ref{fig:potentialD4} compares the scalar effective potential
for the four-dimensional loop-quantum-corrected de Sitter black hole
with those of the corresponding Schwarzschild--de Sitter and
Schwarzschild geometries. We set $\ell=1$, $\gamma=0.21$,
$\Lambda=0.03$, and choose the representative nonextremal
configuration $M_{\rm ext}/M=0.9$ for the quantum-corrected
background.

For the two de Sitter geometries, the physically relevant domain is
the static region bounded by the corresponding event and cosmological
horizons. Since the horizon locations are different in the
loop-quantum-corrected and classical geometries, the LQG--dS and
Schwarzschild--de Sitter curves are displayed over their respective
static intervals and therefore do not have identical radial
endpoints. Both potentials vanish at their corresponding horizons.

As is evident from Fig.~\ref{fig:potentialD4}, the loop-quantum
correction modifies both the height and the radial position of the
effective-potential barrier relative to the Schwarzschild--de Sitter
case. For the parameter values shown, the maximum of the LQG--dS
potential is higher than that of the Schwarzschild--de Sitter
potential. This comparison isolates the effect of the loop-quantum
correction within a de Sitter background. The Schwarzschild curve is
included separately as an asymptotically flat reference; its
substantially larger barrier should therefore not be interpreted as
arising solely from the loop-quantum correction, since the
cosmological background is different. Figure~\ref{fig:potentialD4}
thus shows separately the modification associated with the quantum
correction and the effect of replacing the asymptotically flat
background by a de Sitter one.

\begin{figure}[htbp]
\centering
\includegraphics[width=0.78\linewidth]{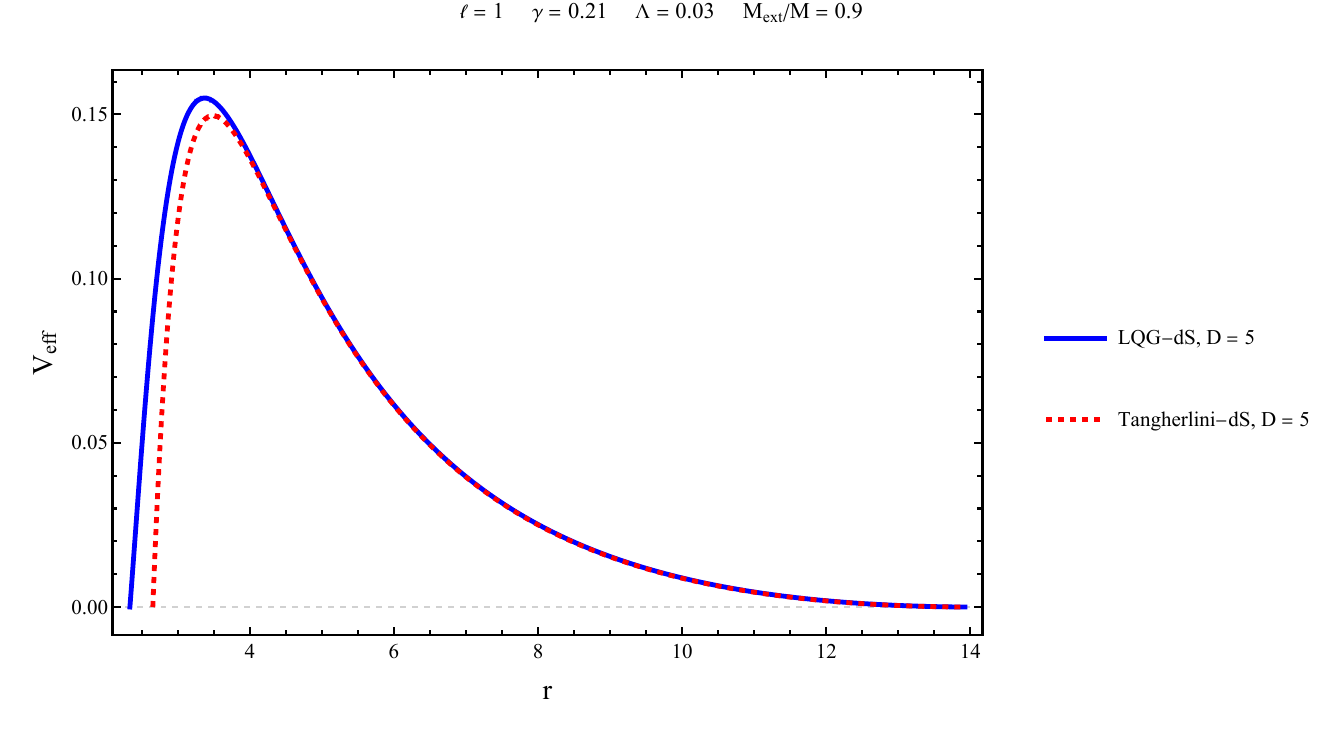}
\caption{\small
Scalar effective potential for the five-dimensional
loop-quantum-corrected de Sitter black hole and the corresponding
Tangherlini--de Sitter background for $\ell=1$,
$\gamma=0.21$, $\Lambda=0.03$, and
$M_{\rm ext}/M=0.9$. The potentials are displayed between their
respective event and cosmological horizons.}
\label{fig:potentialD5}
\end{figure}

Figure~\ref{fig:potentialD5} presents the analogous comparison in
five dimensions, where the classical reference geometry is the
Tangherlini--de Sitter black hole. The quantum-corrected and
classical potentials retain the same single-barrier structure and
vanish at their corresponding event and cosmological horizons.
However, the loop-quantum correction produces a visible modification
near the maximum of the barrier, including a modest change in its
height and radial position. The two potentials approach one another
away from the peak region. This behavior indicates that, for the
chosen parameters, the quantum correction acts primarily in the
strong-field region governing the dominant quasinormal response.

The comparison shows that the deviations of the perturbative
observables from their classical counterparts can be traced directly
to changes in the effective scattering potential. In particular,
loop-quantum corrections modify the barrier in the strong-field
region, while increasing the spacetime dimensionality produces a
much larger change in its overall scale. Since the real part of a
photon-sphere quasinormal frequency is primarily associated with the
characteristic oscillation scale near the potential maximum, whereas
the imaginary part is sensitive to the curvature of the barrier and
the corresponding leakage of the perturbation, these modifications
provide a direct physical explanation for the changes observed in
the quasinormal spectrum. The single-barrier structure in the
physical region $r_h<r<r_c$ is also consistent with the damped
ringdown behavior obtained from the time-domain evolution.

\subsection{Quasinormal-mode boundary conditions and numerical methods}

Having established the physical black-hole domain and the structure of
the scalar effective potential, we now formulate the QNM
problem. For the nonextremal loop-quantum-corrected de Sitter black
holes considered here, the relevant static region is bounded by the
event horizon $r=r_h$ and the cosmological horizon $r=r_c$. The
QNM boundary conditions therefore correspond to purely
ingoing waves at the event horizon and purely outgoing waves at the
cosmological horizon. In terms of the tortoise coordinate, these
conditions are
\begin{equation}
\psi(r)\sim e^{-i\omega r_*},
\qquad r\rightarrow r_h,
\end{equation}
and
\begin{equation}
\psi(r)\sim e^{+i\omega r_*},
\qquad r\rightarrow r_c.
\end{equation}
The resulting boundary-value problem admits a discrete spectrum of
complex frequencies,
\begin{equation}
\omega=\omega_R+i\omega_I,
\end{equation}
where $\omega_R$ determines the oscillation frequency and
$\omega_I<0$ corresponds to a damped perturbation.

In what follows, we compute the lowest-lying QNMs using three complementary numerical methods to ensure accuracy and consistency \cite{ferrari1984new,leaver1985analytic,cho2012new}. We begin with a time-domain integration based on a finite-difference evolution \cite{gundlach1994late} in double-null coordinates, and then extract the quasinormal spectrum using a multi-mode Prony analysis \cite{berti2007mining}. To validate the results, we additionally employ a frequency-domain matrix (determinant) method \cite{lin2017matrix} and the WKB approximation \cite{schutz1985black,kokkotas1988black,iyer1987black}. Introducing double-null coordinates $u=t-r_{\ast}$ and $v=t+r_{\ast}$, the wave equation may be written as
\begin{equation}
-4\,\frac{\partial^{2}\psi}{\partial u\,\partial v}=V_{\mathrm{eff}}(r(u,v))\,\psi.
\end{equation}

Using a standard finite-difference stencil on the null grid, the field value at the point $N=(u+\Delta,v+\Delta)$ can be computed from the neighboring points $W=(u,v+\Delta)$, $E=(u+\Delta,v)$, and $S=(u,v)$ via
\begin{equation}
\psi_{N}=\psi_{W}+\psi_{E}-\psi_{S}-\frac{\Delta u\,\Delta v}{8}\,V_{\mathrm{eff}}(r(u,v))(\psi_{W}+\psi_{E}),
\label{eq:finite_difference}
\end{equation}
where $\Delta u=\Delta v=\Delta$ is the grid spacing. The time-domain profile is obtained once initial data are specified; we choose a Gaussian pulse on one null surface,
\begin{equation}
\psi(u,0)=0,\qquad
\psi(0,v)=\exp\!\left[-\frac{(v-v_{c})^{2}}{2\sigma^{2}}\right],
\end{equation}
with $v_{c}$ and $\sigma$ controlling the center and width of the pulse. The resulting signal $\psi(t,r_{\ast})$ at a fixed extraction point exhibits an oscillatory decay after the prompt response. To extract the quasinormal frequencies, we use a multi-mode Prony fit of the early ringdown waveform to a sum of damped exponentials,
\begin{equation}
\psi(t)=\sum_{j=1}^{p}C_{j}e^{-i\omega_{j}t},
\end{equation}
with $p=2$--$4$ modes in practice, which improves stability in the presence of mode mixing and enables a clean isolation of the fundamental photon-sphere branch. Sampling the signal at times $t=t_{0}+nh$, one may rewrite the data as
\begin{equation}
x_{n}=\psi(t_{0}+nh)=\sum_{j=1}^{p}\tilde{C}_{j}z_{j}^{\,n},\qquad
z_{j}=e^{-i\omega_{j}h},\qquad
\tilde{C}_{j}=C_{j}e^{-i\omega_{j}t_{0}}.
\end{equation}

Introducing the polynomial
\begin{equation}
A(z)=\prod_{j=1}^{p}(z-z_{j})=\sum_{i=0}^{p}\alpha_{i}z^{p-i},\qquad \alpha_{0}=1,
\end{equation}
the roots $z_{j}$ follow from a linear system constructed from the sampled data, and the quasinormal frequencies are recovered by
\begin{equation}
\omega_{j}=\frac{i}{h}\ln z_{j}.
\end{equation}

As an independent cross-check, we also compute the spectrum using the matrix method. Near a simple horizon $r=r_{\pm}$, the metric behaves as $f(r)\simeq f'(r_{\pm})(r-r_{\pm})$, and the quasinormal boundary behaviors can be written as
\begin{equation}
\psi(r)\sim (r-r_{h})^{-i\omega/f'(r_{h})},\qquad
\psi(r)\sim (r_{c}-r)^{+i\omega/f'(r_{c})}.
\end{equation}

Mapping the domain $r\in[r_{h},r_{c}]$ to $y\in[0,1]$ using $y=(r-r_{h})/(r_{c}-r_{h})$, we factor out the known horizon behavior and write the wavefunction in the form
\begin{equation}
\psi(r)=y^{-i\omega/f'(r_{h})}(1-y)^{+i\omega/f'(r_{c})}\,Y(y),
\end{equation}
so that $Y(y)$ is regular on the compact interval. The radial equation becomes a second-order ordinary differential equation of the generic form
\begin{equation}
b_{0}(\omega,y)Y(y)+b_{1}(\omega,y)Y'(y)+b_{2}(\omega,y)Y''(y)=0,
\end{equation}
supplemented by regular boundary conditions at $y=0$ and $y=1$. Discretizing the equation yields a matrix system $\Gamma(\omega)Y=0$, and QNMs are obtained from the nonlinear condition $\det\Gamma(\omega)=0$. Finally, we employ the WKB approximation as a semi-analytic method. The six-order WKB formula can be written in the standard form
\begin{equation}
\frac{i}{\sqrt{-2V_{\mathrm{eff}}''(r_{0})}}
\left[\omega^{2}-V_{\mathrm{eff}}(r_{0})+\sum_{j=2}^{6}\Lambda_{j}\right]
=n+\frac{1}{2},
\end{equation}
where primes denote derivatives with respect to the tortoise coordinate $r_{\ast}$, $r_{0}$ is the location of the maximum of the effective potential, $n$ is the overtone number, and the correction terms $\Lambda_{j}$ are the higher-order WKB contributions available in the standard literature.

\subsection{Time-domain evolution and quasinormal-mode spectrum}
\label{subsec:qnm_results}

\begin{figure*}[htbp]
\centering
\includegraphics[width=0.96\textwidth]
{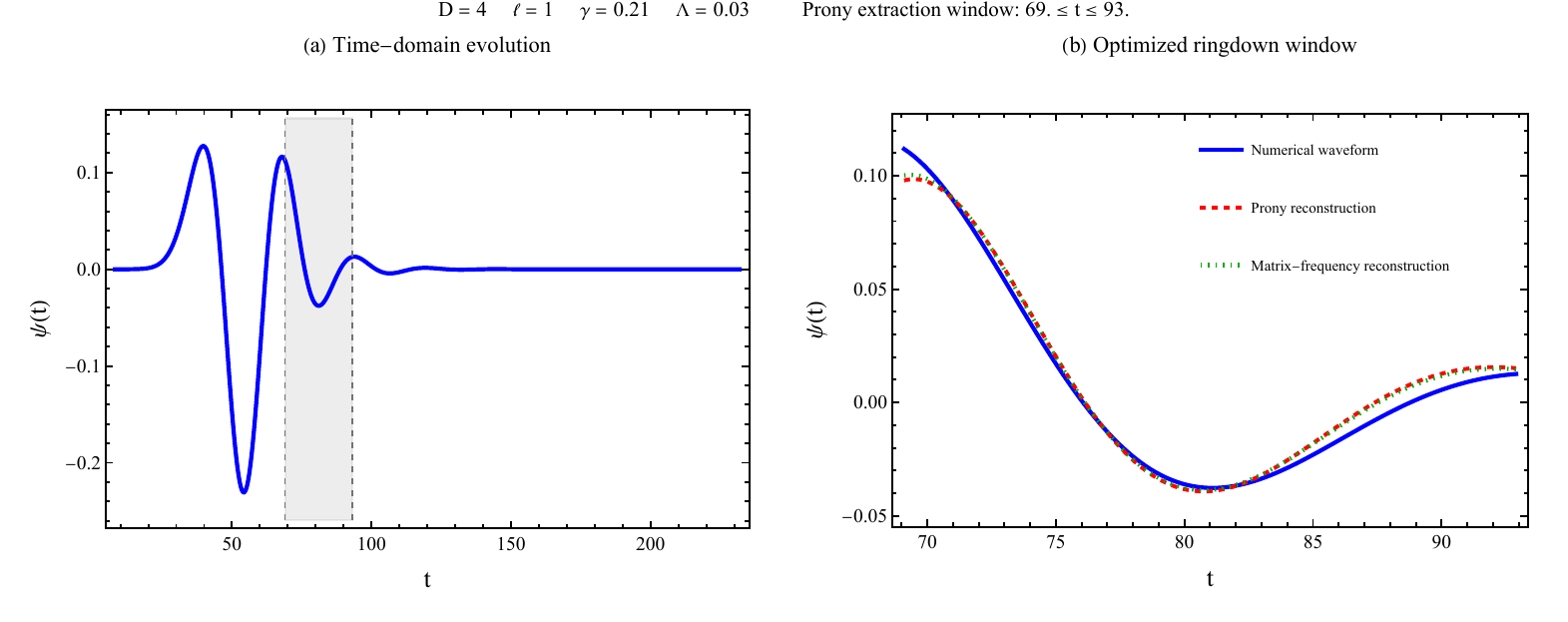}
\caption{\small
Time-domain evolution and QNM reconstruction of the massless scalar
perturbation for the representative $D=4$ configuration with
$\ell=1$, $\gamma=0.21$, $\Lambda=0.03$, and
$M_{\rm ext}/M=0.9$.
Panel (a) shows the complete numerical waveform obtained from the
finite-difference evolution. The shaded region,
$69\lesssim t\lesssim93$, identifies the ringdown interval whose
numerical samples are used as input to the multi-mode Prony extraction;
the shaded region itself is not a fitted curve.
Panel (b) magnifies the same interval and compares the numerical
waveform with the Prony reconstruction and a reconstruction based on
the independently determined matrix-method frequency. The agreement
in oscillation phase and damping provides a direct graphical
consistency check of the extracted fundamental QNM frequency. The
corresponding numerical frequencies are listed in
Table~\ref{tab:tableI}.
}
\label{fig:td_ringdown}
\end{figure*}

To provide a direct graphical assessment of the time-domain calculation
and of the subsequent extraction of the quasinormal frequencies, we
show in Fig.~\ref{fig:td_ringdown} a representative evolution of the
massless scalar perturbation together with its ringdown reconstruction.
We consider the fundamental scalar mode with $\ell=1$ for the
four-dimensional loop-quantum-corrected de Sitter black hole, taking
$\gamma=0.21$ and $\Lambda=0.03$. As in the remainder of the QNM
analysis, the mass is chosen inside the admissible nonextremal interval,
$M_{\rm ext}<M<M_{\rm N}$, so that the evolution is performed in the
static region bounded by the event and cosmological horizons.

Figure~\ref{fig:td_ringdown}(a) displays the complete numerical waveform
obtained by solving Eq.~(33) with the finite-difference evolution
described above. The signal exhibits the standard sequence expected for
black-hole perturbations: an initial response determined by the
prescribed initial data, followed by a damped oscillatory ringdown.
The latter is the interval in which the response is dominated by the
lowest-lying QNMs and is therefore appropriate for the extraction of
the complex frequencies. The shaded region in panel (a) does not
represent a fit by itself; rather, it identifies the time interval
selected for the subsequent Prony extraction.

To avoid contamination from the initial transient and the late-time
portion of the signal, we scan candidate ringdown windows and select
the interval over which the damped-mode reconstruction provides the
most stable description of the numerical waveform. For the
representative configuration shown in Fig.~\ref{fig:td_ringdown}, this
procedure gives approximately $69\lesssim t\lesssim93$. The numerical
samples $\psi(t_n)$ within this interval are then used as input to the
multi-mode Prony procedure defined in Eqs.~(35)--(38). The signal is
represented as a superposition of damped complex exponentials, and the
corresponding Prony roots $z_j$ determine the complex QNM frequencies
through Eq.~(38). Thus, the QNM frequency is not read directly from
panel (a); panel (a) identifies the portion of the numerical waveform
used for the frequency extraction.

Figure~\ref{fig:td_ringdown}(b) provides the corresponding graphical
verification of the extraction. It magnifies the same ringdown interval
indicated in panel (a) and compares the numerical waveform with the
Prony reconstruction obtained from the extracted complex frequency.
For an independent cross-check, we also show a reconstruction based on
the QNM frequency determined by the matrix method, with the corresponding
amplitude and phase fitted to the same numerical waveform. Both
reconstructions closely reproduce the oscillation phase and exponential
damping of the numerical signal throughout the selected interval. This
agreement provides a direct time-domain consistency check that the Prony
and matrix calculations identify the same dominant quasinormal response.
The corresponding Prony, matrix-method, and WKB frequencies for this
representative configuration are reported in Table~\ref{tab:tableI},
together with the results for the other values of the loop-quantum
parameter considered below.

We now extend this representative time-domain comparison to the
quasinormal-mode spectrum over the parameter space considered in this
work. For each set of $(D,\gamma,\Lambda)$, the mass is chosen within
the admissible three-horizon interval established in
Sec.~\ref{subsec:horizon_structure}. This guarantees that the QNM
boundary conditions at the event and cosmological horizons are
well defined for all configurations considered below.
Table~\ref{tab:tableI} lists the quasinormal frequencies for $D=4,5$
and several values of the loop-quantum parameter $\gamma$, while
Table~\ref{tab:tableII} shows their dependence on the spacetime
dimension for fixed $\gamma=0.21$. In all cases, the three approaches
agree at the level of a few percent, providing a useful cross-check
for the low-multipole fundamental modes considered here.

\begin{table}[htbp]
\centering
\caption{Lowest-lying quasinormal modes for $\ell=1$ and varying
$\gamma$ ($\Lambda=0.03$).}
\label{tab:tableI}
\renewcommand{\arraystretch}{1.25}
\begin{tabular}{|c|c|c|c|c|}
\hline
$D$ & $\gamma$ & $\omega$ (Prony) & $\omega$ (Matrix methods)
& $\omega$ (WKB approximation) \\
\hline
4 & 0.09 & $0.2759064-0.0831822i$ & $0.2723446-0.0850307i$
& $0.2739885-0.0840224i$ \\
\hline
4 & 0.18 & $0.2775052-0.0822443i$ & $0.2739227-0.0840719i$
& $0.2755762-0.0830751i$ \\
\hline
4 & 0.21 & $0.2786625-0.0817254i$ & $0.2750650-0.0835415i$
& $0.2767254-0.0825509i$ \\
\hline
5 & 0.09 & $1.2334262-0.4555248i$ & $1.2175031-0.4656476i$
& $1.2248522-0.4601261i$ \\
\hline
5 & 0.18 & $1.1625328-0.5358541i$ & $1.1475249-0.5477619i$
& $1.1544516-0.5412668i$ \\
\hline
5 & 0.21 & $1.1198099-0.5290735i$ & $1.1053535-0.5408307i$
& $1.1120257-0.5344177i$ \\
\hline
\end{tabular}
\end{table}

\begin{table}[htbp]
\centering
\caption{Dependence of the fundamental mode on spacetime dimension
for fixed $\gamma=0.21$ ($\ell=1$, $\Lambda=0.03$).}
\label{tab:tableII}
\renewcommand{\arraystretch}{1.25}
\begin{tabular}{|c|c|c|c|}
\hline
$D$ & $\omega$ (Prony) & $\omega$ (Matrix methods)
& $\omega$ (WKB approximation) \\
\hline
4 & $0.2786625-0.0817254i$ & $0.2750650-0.0835415i$
& $0.2767254-0.0825509i$ \\
\hline
5 & $1.1198099-0.5290735i$ & $1.1053535-0.5408307i$
& $1.1120257-0.5344177i$ \\
\hline
6 & $2.1539476-1.1407175i$ & $2.1261419-1.1660668i$
& $2.1389748-1.1522399i$ \\
\hline
7 & $3.1620245-1.7656172i$ & $3.1221900-1.8048531i$
& $3.1410372-1.7834517i$ \\
\hline
8 & $4.1533260-2.3894211i$ & $4.0997091-2.4425194i$
& $4.1244558-2.4135567i$ \\
\hline
\end{tabular}
\end{table}

Several noteworthy features follow from
Tables~\ref{tab:tableI} and~\ref{tab:tableII}. First, the multi-mode
Prony time-domain extraction, the matrix determinant method, and the
WKB approximation produce mutually consistent spectra, confirming the
reliability of the numerical implementation and the stability of the
extracted mode family. In particular, the representative
$D=4$, $\gamma=0.21$ frequencies used in the reconstruction shown in
Fig.~\ref{fig:td_ringdown} are given directly in
Table~\ref{tab:tableI}, thereby allowing the graphical waveform
comparison and the numerical frequency comparison to be read together.

Second, the loop parameter $\gamma$ introduces controlled quantitative
shifts in both $\omega_R$ and $\omega_I$, reflecting the modification
of the effective-potential barrier through the metric function $f(r)$.
In four dimensions, $\omega_R$ increases mildly while the magnitude of
the damping rate decreases slightly as $\gamma$ grows. In five
dimensions, by contrast, the oscillation frequency decreases and the
magnitude of the damping rate generally increases over the range
considered. The response of the quasinormal spectrum to the
loop-quantum correction is therefore dimension dependent.

The dependence on spacetime dimensionality is considerably stronger.
As shown in Table~\ref{tab:tableII}, for fixed $\gamma=0.21$ both the
oscillation frequency $\omega_R$ and the damping rate $|\omega_I|$
increase rapidly with $D$. This behavior is consistent with the
corresponding modification of the effective scattering barrier and
the photon-sphere dynamics in higher dimensions. Such dimensional
dependence is a generic feature of photon-sphere QNMs in
higher-dimensional black holes and is not peculiar to the
loop-quantum correction; see, for example,
Ref.~\cite{konoplya2011quasinormal} and references therein.

An important outcome of our analysis is that all computed quasinormal
frequencies possess negative imaginary parts. Within the parameter
range investigated here, this indicates the absence of exponentially
growing scalar modes and therefore supports the dynamical stability of
the loop-quantum-corrected black holes against the class of massless
scalar perturbations considered. The consistency among the three
computational approaches, together with the systematic dependence of
the spectrum on $\gamma$ and $D$, further indicates that the
loop-quantum correction produces quantitative shifts of the
quasinormal spectrum while preserving the damped photon-sphere
ringdown behavior.

\begin{figure}[htb]
\centering
\begin{tabular}{cc}
\includegraphics[width=0.46\textwidth]{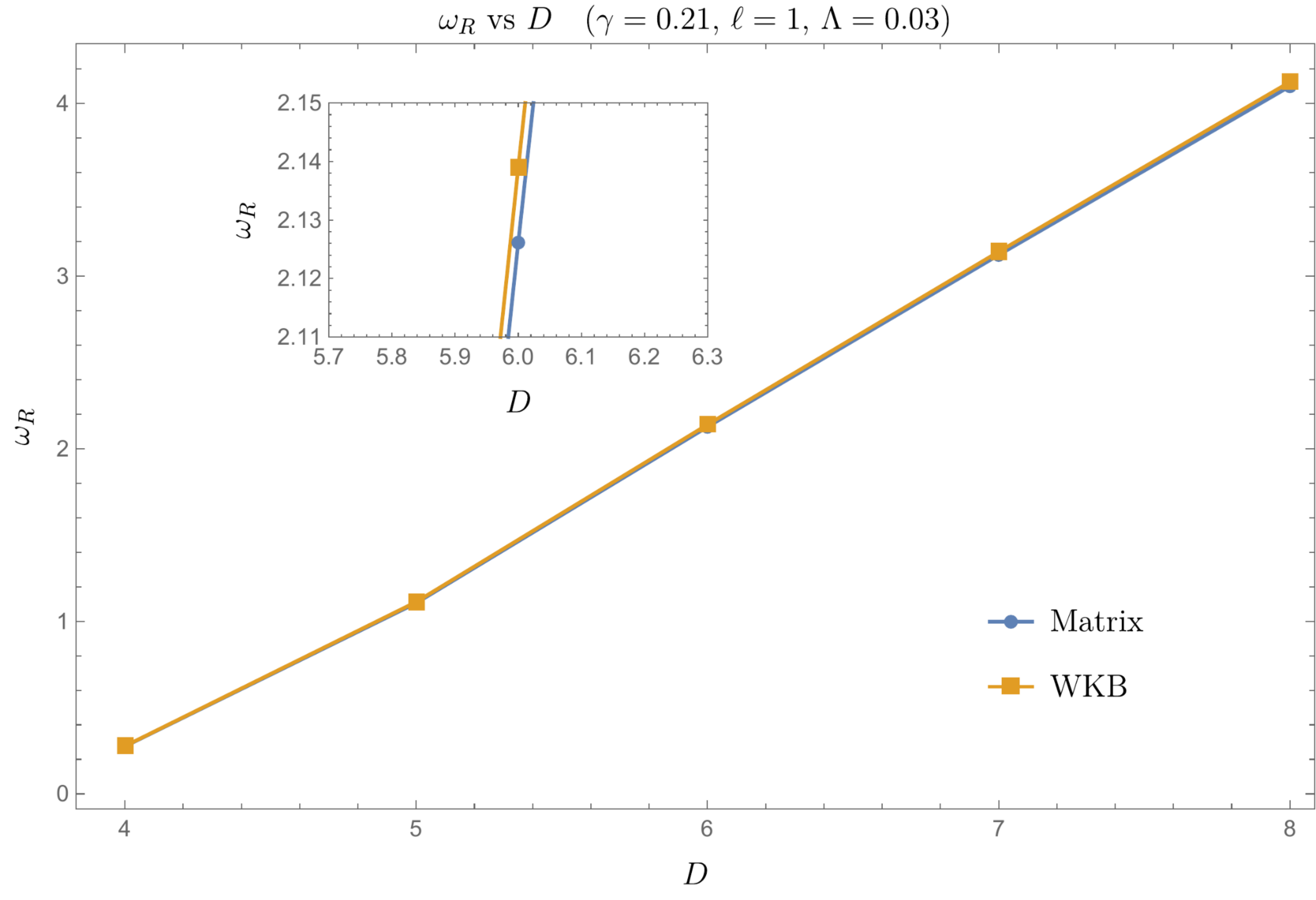}\, &
\includegraphics[width=0.455\textwidth]{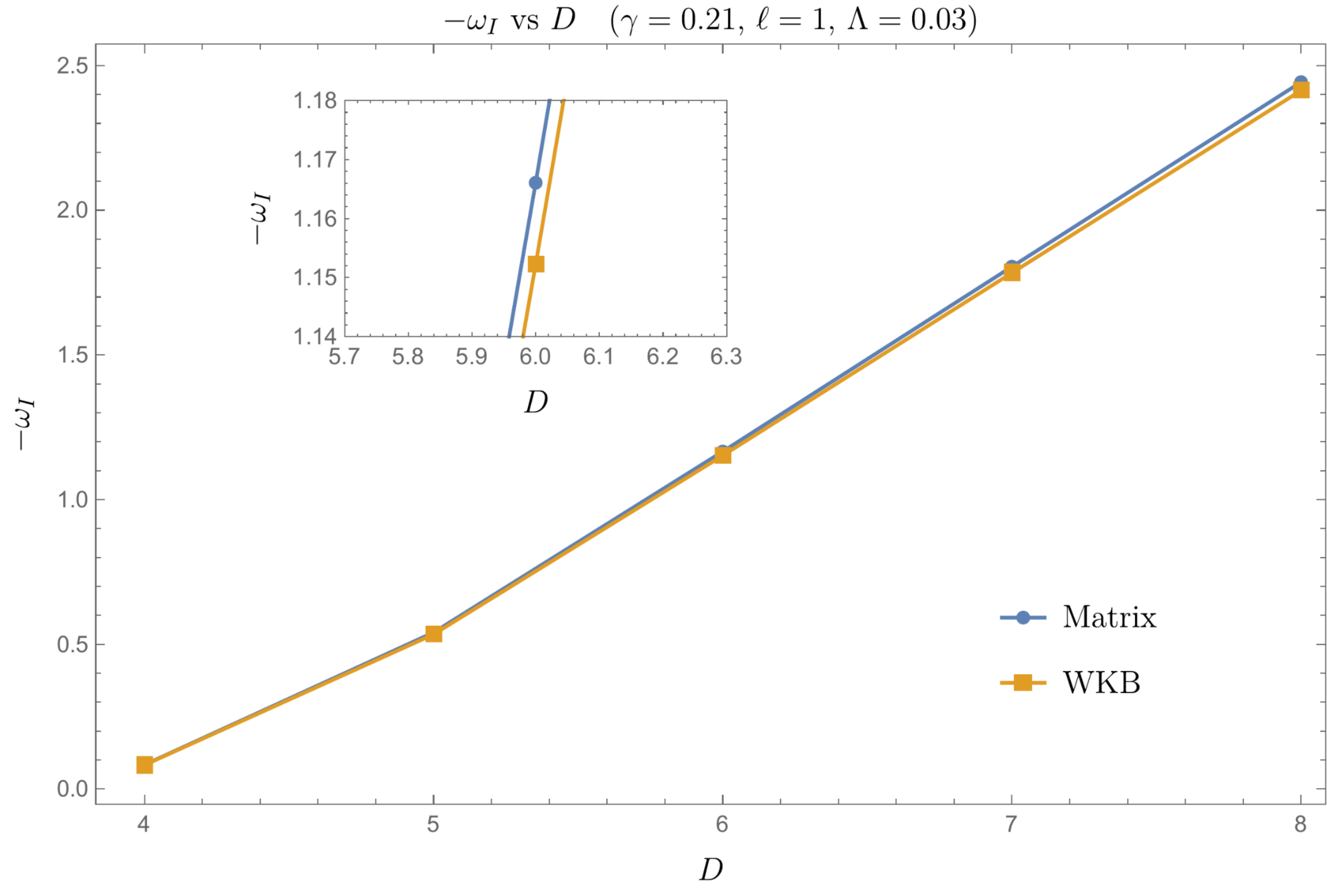}
\end{tabular}
\caption{\label{Figgt}\small{\emph{
Real and imaginary parts of the fundamental scalar quasinormal
frequency as functions of the spacetime dimension $D$ for fixed
$\gamma=0.21$, $\ell=1$, and $\Lambda=0.03$. The results obtained
using the matrix method and the WKB approximation are shown. The inset
magnifies the region around $D\simeq6$, where the two methods are very
close but remain distinguishable.}}}
\end{figure}

\begin{figure}[htb]
\centering
\begin{tabular}{cc}
\includegraphics[width=0.46\textwidth]{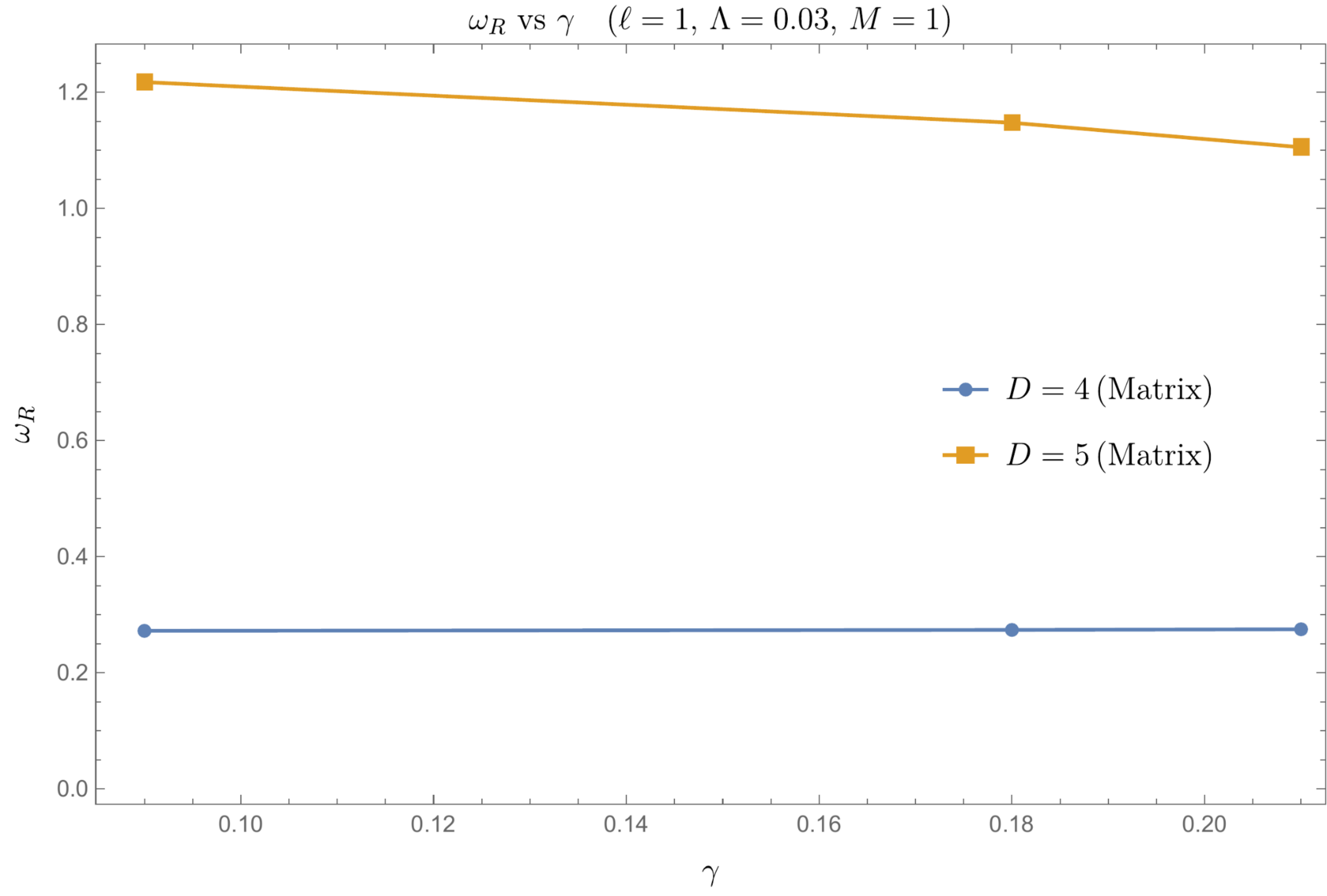}\, &
\includegraphics[width=0.455\textwidth]{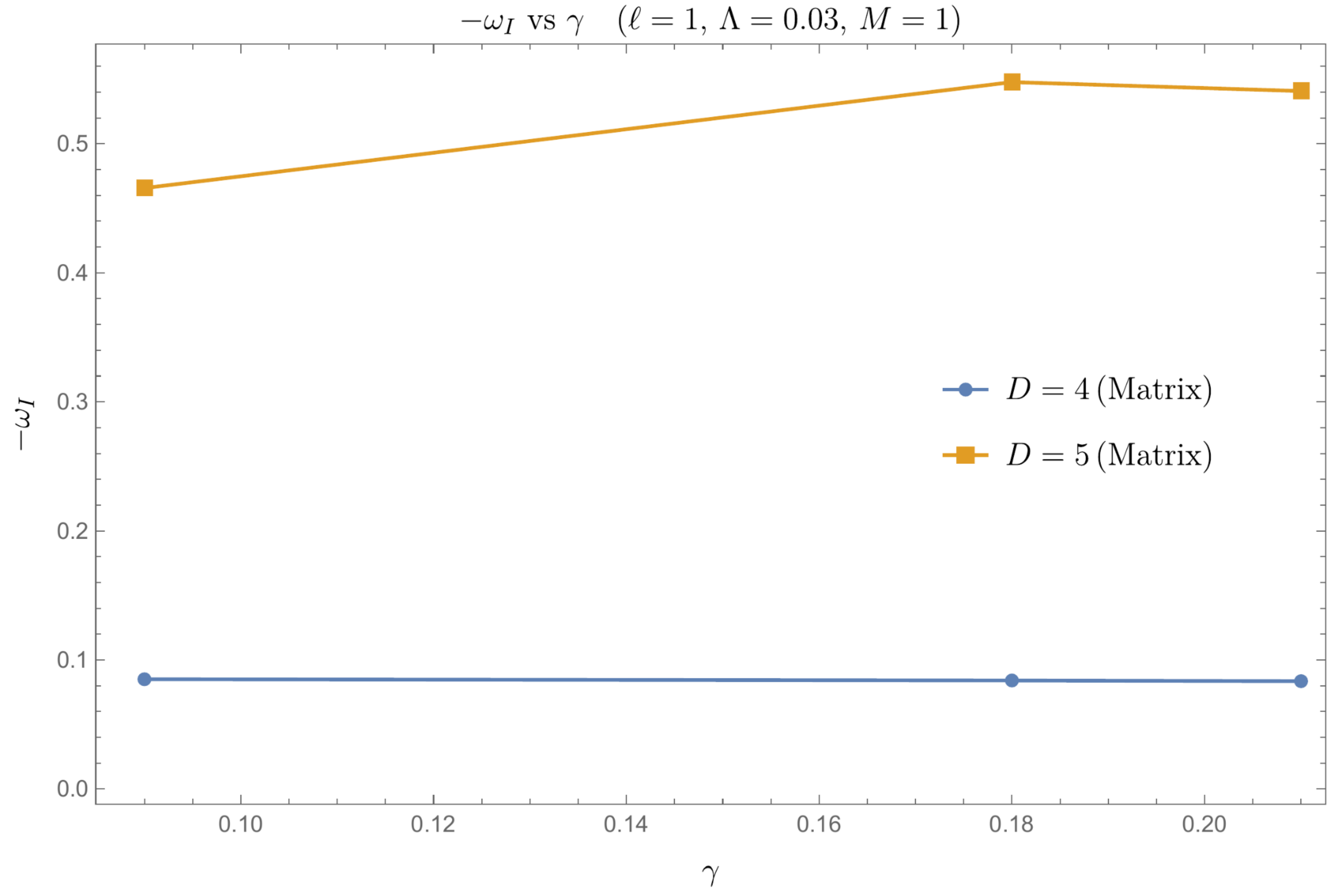}
\end{tabular}
\caption{\label{Figgt2}\small{\emph{
Dependence of the real and imaginary parts of the fundamental scalar
quasinormal frequency on the loop-quantum parameter $\gamma$ for
$D=4$ and $D=5$, with $\ell=1$ and $\Lambda=0.03$. All frequencies
are obtained using the matrix method.}}}
\end{figure}

To further illustrate the numerical results presented in
Tables~\ref{tab:tableI} and~\ref{tab:tableII}, we display the behavior
of the quasinormal frequencies graphically in
Figs.~\ref{Figgt} and~\ref{Figgt2}. Figure~\ref{Figgt} shows the
dependence of the real and imaginary parts of the fundamental scalar
quasinormal mode on the spacetime dimension $D$ for fixed
$\gamma=0.21$. Both $\omega_R$ and $|\omega_I|$ exhibit a pronounced
growth as the number of dimensions increases. The matrix and WKB
results are in close agreement, and the inset highlights their small
quantitative differences around $D\simeq6$.

Figure~\ref{Figgt2} displays the dependence of the same mode on the
loop-quantum parameter $\gamma$ for $D=4$ and $D=5$. While the effect
of $\gamma$ is relatively mild in four dimensions, the
higher-dimensional case exhibits a more noticeable response,
particularly in the damping rate. These plots provide a visual
representation of the trends reported in the numerical tables and
show that the influence of spacetime dimensionality is substantially
stronger than the quantitative shifts produced by the loop-quantum
parameter.

\noindent\textbf{Comparison with the classical Schwarzschild limit.}\quad
To make the connection with the standard general-relativistic result
explicit, we finally compare the fundamental scalar QNM of the
four-dimensional loop-quantum-corrected de Sitter black hole with its
classical counterparts. From Eq.~(11), the Schwarzschild--de Sitter
geometry is recovered by switching off the loop-quantum correction,
$\alpha\rightarrow0$, while keeping $\Lambda$ fixed. Taking in addition
the asymptotically flat limit $\Lambda\rightarrow0$ gives
\begin{equation}
f(r)\longrightarrow f_{\rm Sch}(r)
=1-\frac{2GM}{r},
\end{equation}
which is the standard Schwarzschild geometry.

This comparison is useful because the two successive limits separate
the physical origins of the deviations in the QNM spectrum. The
comparison between the LQG--dS and Schwarzschild--de Sitter cases,
performed at the same mass, multipole number, and cosmological
constant, isolates the modification associated with the loop-quantum
correction. The subsequent comparison with Schwarzschild additionally
shows the effect of removing the de Sitter background.

\begin{figure*}[htbp]
\centering
\includegraphics[width=0.96\textwidth]
{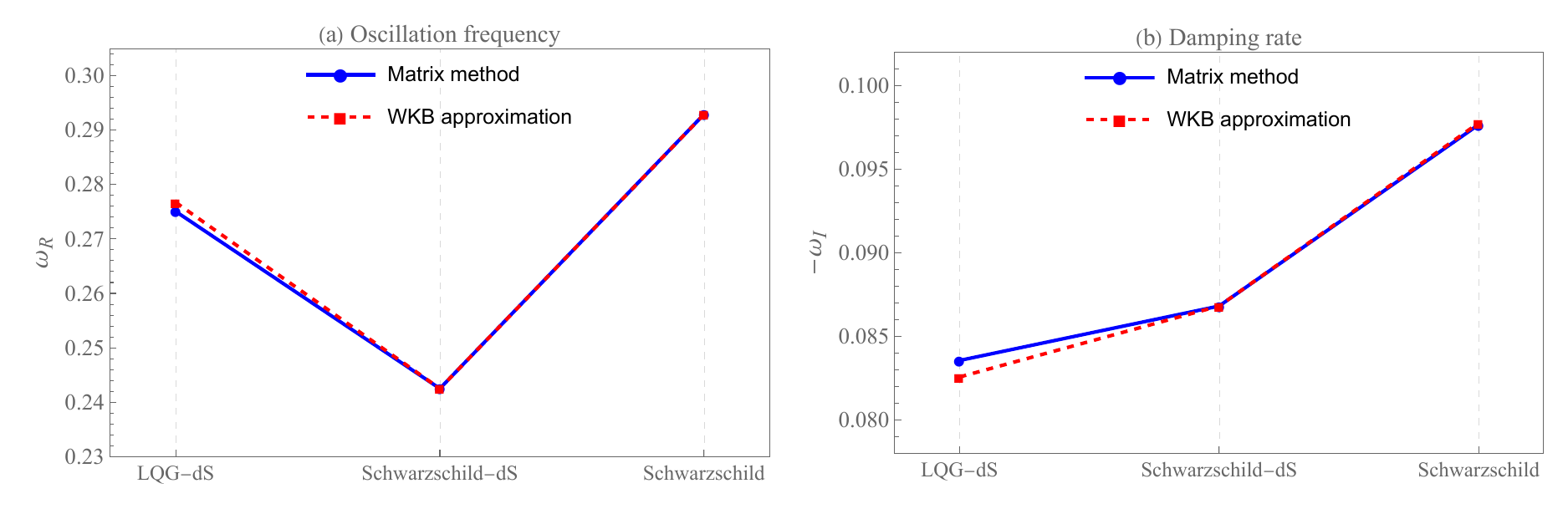}
\caption{\small
Comparison of the fundamental scalar QNM for the
four-dimensional loop-quantum-corrected de Sitter black hole and its
classical limits for $\ell=1$. Panel (a) shows the oscillation
frequency $\omega_R$, while panel (b) shows the damping rate
$-\omega_I$. The Schwarzschild--de Sitter result is obtained by
switching off the loop-quantum correction, $\alpha=0$, at fixed
$\Lambda=0.03$, whereas the standard Schwarzschild result follows
from the simultaneous limit $\alpha=0$ and $\Lambda=0$. The
comparison separates the contribution of the loop-quantum deformation
from that associated with the cosmological background.}
\label{fig:qnm_classical_comparison}
\end{figure*}

Table~\ref{tab:qnm_classical_comparison} summarizes the fundamental
$\ell=1$ scalar QNM frequencies for these three geometries. The
corresponding comparison is displayed graphically in
Fig.~\ref{fig:qnm_classical_comparison}, where the real part
$\omega_R$ and the damping rate $-\omega_I$ are shown separately.
The Schwarzschild value is obtained by setting
$\alpha=0$ and $\Lambda=0$ in the four-dimensional metric and solving
the same scalar perturbation problem with the appropriate
asymptotically flat QNM boundary condition at spatial infinity.

\begin{table}[htbp]
\centering
\caption{Comparison of the fundamental scalar QNM
($\ell=1$, $D=4$) for the loop-quantum-corrected de Sitter,
Schwarzschild--de Sitter, and Schwarzschild geometries. The
Schwarzschild limit is obtained by setting
$\alpha=0$ and $\Lambda=0$.}
\label{tab:qnm_classical_comparison}
\begin{tabular}{lccc}
\hline\hline
Geometry & Parameters & $\omega_{\rm Matrix}$ & $\omega_{\rm WKB}$ \\
\hline
LQG--dS
& $\gamma=0.21,\ \Lambda=0.03$
& $0.2750650-0.0835415\,i$
& $0.2767254-0.0825509\,i$ \\

Schwarzschild--dS
& $\alpha=0,\ \Lambda=0.03$
& $0.2425-0.0868\,i$
& $0.2425-0.0868\,i$ \\

Schwarzschild
& $\alpha=0,\ \Lambda=0$
& $0.292936-0.09766\,i$
& $0.292910-0.097762\,i$ \\
\hline\hline
\end{tabular}
\end{table}

The comparison shows explicitly how the quantum-corrected de Sitter
spectrum approaches the standard classical result as the deformation
parameters are removed. In particular, the loop-quantum correction
produces a quantitative shift of both the oscillation frequency and
the damping rate rather than introducing a qualitatively distinct
mode family. This behavior is consistent with the effective-potential
analysis of Sec.~III\,A: the deformation modifies the height and
curvature of the photon-sphere potential barrier, while the dominant
ringdown remains associated with the same photon-sphere branch.
The Schwarzschild comparison therefore provides a useful classical
benchmark for assessing the magnitude of the modifications generated
by the loop-quantum and cosmological terms.

\section{Shadow Signatures}

\subsection{Null Geodesics Formalism}\label{se3}

In order to investigate the optical appearance of the
higher-dimensional loop quantum de Sitter black hole, we analyze the
behavior of null geodesics in this spacetime. The shadow of a black
hole is determined by photon trajectories that asymptotically approach
unstable circular orbits, commonly referred to as the photon sphere.
The role of photon-sphere dynamics in higher-dimensional Tangherlini
spacetimes and its connection with strong gravitational lensing has
been studied extensively; see, for example,
Ref.~\cite{Tsukamoto2014}. Consequently, the study of null geodesics
provides direct insight into the geometric structure governing both
optical observables and high-frequency perturbations.

The motion of a test particle is described by the Lagrangian
\begin{equation}
\bar{\mathcal{L}}=\frac{1}{2}g_{\mu \nu}\dot{x}^{\mu} \dot{x}^{\nu},
\end{equation}
where the overdot denotes differentiation with respect to the affine parameter $\tau$. Due to the stationarity and spherical symmetry of the spacetime, there exist conserved quantities associated with time translations and rotations. The corresponding conjugate momenta are given by
\begin{equation}
P_t=h(r)\dot{t}=E, \qquad 
P_{\theta_{D-2}}=L,
\end{equation}
where $E$ and $L$ represent the conserved energy and angular momentum of the photon, respectively.

To derive the equations of motion, we employ the Hamilton--Jacobi formalism. The action $S$ satisfies
\begin{equation}
\frac{\partial S}{\partial \tau}= -\frac{1}{2}g^{\mu \nu} \frac{\partial S}{\partial x^\mu}\frac{\partial S}{\partial x^\nu}.
\end{equation}
Assuming separability of the action in arbitrary dimensions,
\begin{equation}
S=-Et+L\theta_{D-2}+S_r(r)+\sum_{i=1}^{D-3}S_{\theta_i}(\theta_i),
\end{equation}
and restricting to null geodesics, one obtains decoupled radial and angular equations.
The radial motion can be expressed in the compact form
\begin{equation}
r^2 \dot{r}=\pm \sqrt{\mathcal{R}(r)},
\end{equation}
with
\begin{equation}
\mathcal{R}(r)=r^4 \frac{f(r)}{h(r)}E^2 - r^2 f(r)(L^2+\mathcal{K}),
\end{equation}
where $\mathcal{K}$ is the Carter constant. The radial motion can be conveniently recast into the form

\begin{equation}
\dot r^{\,2}+V_{\rm geo}(r)=0,
\end{equation}

where

\begin{equation}
V_{\rm geo}(r)
=
\frac{f(r)}{r^2}\left(L^2+\mathcal{K}\right)
-
\frac{f(r)}{h(r)}E^2
\label{Vgeo}
\end{equation}

is an effective potential governing the null geodesic dynamics.

\noindent
It should be emphasized that $V_{\rm geo}(r)$ characterizes photon
trajectories and should not be confused with the scalar-field
effective potential $V_{\rm eff}(r)$. The photon-sphere radius
$r_{ph}$, corresponding to the unstable circular null orbit, is
determined by the conditions
\begin{equation}
V_{\rm geo}(r_{ph})=0,
\qquad
\left.\frac{dV_{\rm geo}}{dr}\right|_{r=r_{ph}}=0,
\end{equation}
which lead to the compact relation
\begin{equation}
r_{ph}h'(r_{ph})-2h(r_{ph})=0.
\end{equation}

where the prime denotes differentiation with respect to the radial coordinate $r$. This condition encodes the essential geometric structure of the spacetime and plays a central role in both shadow formation and quasinormal mode behavior. In particular, in the eikonal regime, the real and imaginary parts of quasinormal frequencies are governed by the properties of this unstable photon orbit, establishing a direct link between the shadow and the dynamical response analyzed in Sec.~III.

\subsection{Shadow Construction}\label{se4}

The shadow of a black hole corresponds to the set of directions from which photons fail to reach a distant observer, instead being captured by the event horizon after spiraling around unstable circular orbits. To characterize the shadow, it is convenient to introduce the impact parameters
\begin{equation}
\xi=\frac{L}{E}, \qquad \eta=\frac{\mathcal{K}}{E^2}.
\end{equation}
In terms of these quantities, the radial function takes the form
\begin{equation}
\mathcal{R}(r)=E^2\left[r^4\frac{f(r)}{h(r)}-r^2f(r)(\eta+\xi^2)\right].
\end{equation}

Imposing the photon sphere conditions yields
\begin{equation}
\eta+\xi^2=\frac{r_{ph}^2}{2f(r_{ph})+r_{ph} f'(r_{ph})}
\left[
4\frac{f(r_{ph})}{h(r_{ph})}+
r_{ph}\frac{f'(r_{ph})h(r_{ph})-f(r_{ph})h'(r_{ph})}{h^2(r_{ph})}
\right].
\end{equation}

To describe the shadow as observed at infinity, we introduce the
celestial coordinates \cite{vazquez2003strong}, where $r_{O}$ denotes the radial
coordinate of the observer: 
\begin{equation}
\lambda=\lim_{r_O\to\infty}\left(\frac{r_O^2 P^{(\theta_{D-2})}}{P^t}\right),
\qquad
\psi=\lim_{r_O\to\infty}\left(\frac{r_O^2 P^{(\theta_i)}}{P^t}\right).
\end{equation}
On the equatorial plane, these reduce to
\begin{equation}
\lambda=-\xi, \qquad \psi=\pm\sqrt{\eta},
\end{equation}
leading to the shadow radius
\begin{equation}
R_{Sh}^2=\lambda^2+\psi^2=\eta+\xi^2.
\end{equation}

The numerical values of the photon sphere radius $r_{ph}$ and shadow radius $R_{Sh}$ are summarized in Tables~\ref{tab:shadow_gamma_D4} and \ref{tab:shadow_lambda_dimension}. These results show that loop quantum corrections systematically modify the photon sphere structure. In particular, increasing the parameter $\gamma$ reduces both $r_{ph}$ and $R_{Sh}$, indicating that quantum effects effectively shrink the region of unstable photon motion. On the other hand, the cosmological constant $\Lambda$ acts to enlarge the shadow size by pushing the photon sphere outward. The influence of spacetime dimensionality is even more pronounced, as higher dimensions significantly reduce the shadow radius due to the inward shift of the photon sphere

\begin{table*}[htb]
\centering
\caption{
Photon-sphere radius $r_{ph}$ and shadow radius $R_{\rm Sh}$ for the four-dimensional loop quantum de Sitter black hole for different values of the loop parameter $\gamma$. We set $\Lambda=0.02$.
}
\label{tab:shadow_gamma_D4}
\resizebox{0.26\textwidth}{!}{
\begin{tabular}{c|cc}
\hline
 & \multicolumn{2}{c}{$D=4$} \\
\cline{2-3}
$\gamma$ & $r_{ph}$ & $R_{\rm Sh}$ \\
\hline
0.15   & 2.998330 & 5.19398 \\
0.18   & 2.957780 & 5.14139 \\
0.21   & 2.823620 & 4.96856 \\
0.2375 & 2.566310 & 4.64327 \\
0.25   & 2.056000 & 4.03881 \\
\hline
\end{tabular}
}
\end{table*}

\begin{table*}[htb]
\centering
\caption{
Photon-sphere radius $r_{ph}$ and shadow radius $R_{\rm Sh}$ for different spacetime dimensions and different values of the cosmological constant. We set $\gamma=0.18$.
}
\label{tab:shadow_lambda_dimension}
\resizebox{\textwidth}{!}{
\begin{tabular}{c|cc|cc|cc|cc|cc}
\hline
 & \multicolumn{2}{c|}{$\Lambda=0.02$}
 & \multicolumn{2}{c|}{$\Lambda=0.04$}
 & \multicolumn{2}{c|}{$\Lambda=0.06$}
 & \multicolumn{2}{c|}{$\Lambda=0.08$}
 & \multicolumn{2}{c}{$\Lambda=0.10$} \\
\cline{2-11}
$D$ 
& $r_{ph}$ & $R_{\rm Sh}$ 
& $r_{ph}$ & $R_{\rm Sh}$ 
& $r_{ph}$ & $R_{\rm Sh}$ 
& $r_{ph}$ & $R_{\rm Sh}$ 
& $r_{ph}$ & $R_{\rm Sh}$ \\
\hline
$D=4$ 
& 2.98973 & 4.83160
& 9.43364 & 11.26730
& -- & --
& -- & --
& -- & -- \\

$D=5$ 
& 1.30755 & 1.85798
& 1.60535 & 3.39358
& 1.85594 & 5.15226
& 2.07696 & 7.05714
& 2.27768 & 9.04479 \\

$D=7$ 
& 0.995441 & 1.23020
& 1.10263 & 2.04775
& 1.18537 & 2.93479
& 1.25370 & 3.87576
& 1.31240 & 4.86063 \\
\hline
\end{tabular}
}
\end{table*}

\subsection{Shadow Geometry}

\begin{figure}[htb] 
\centering
 \includegraphics[width=0.4\textwidth]{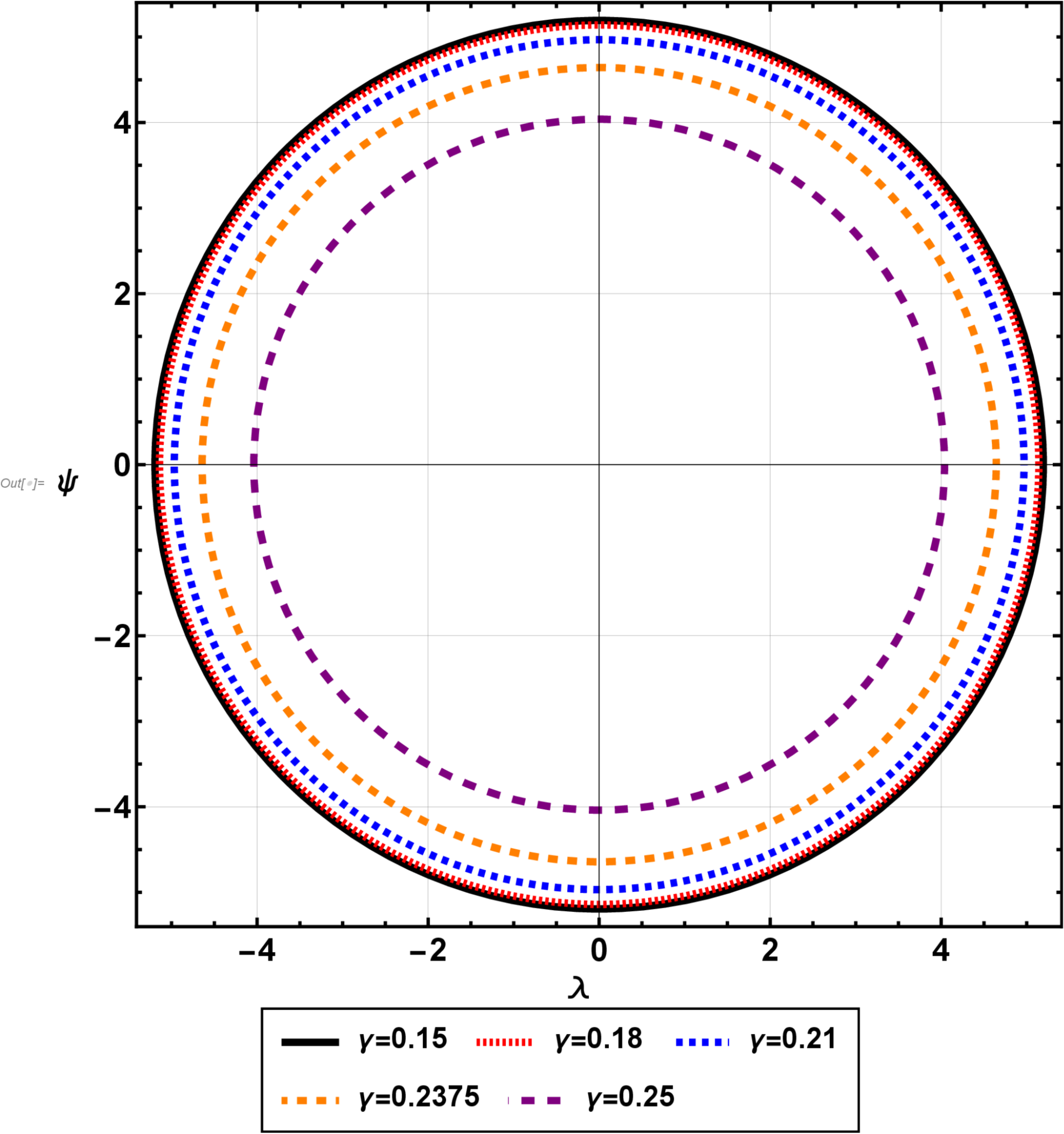} 
 \caption{\small Shadow of the four-dimensional loop quantum de Sitter black hole in celestial coordinates for different values of $\gamma$ and $\Lambda$.} \label{figure4}
  \end{figure}

\begin{figure}[htb] 
\centering 
\includegraphics[width=0.4\textwidth]{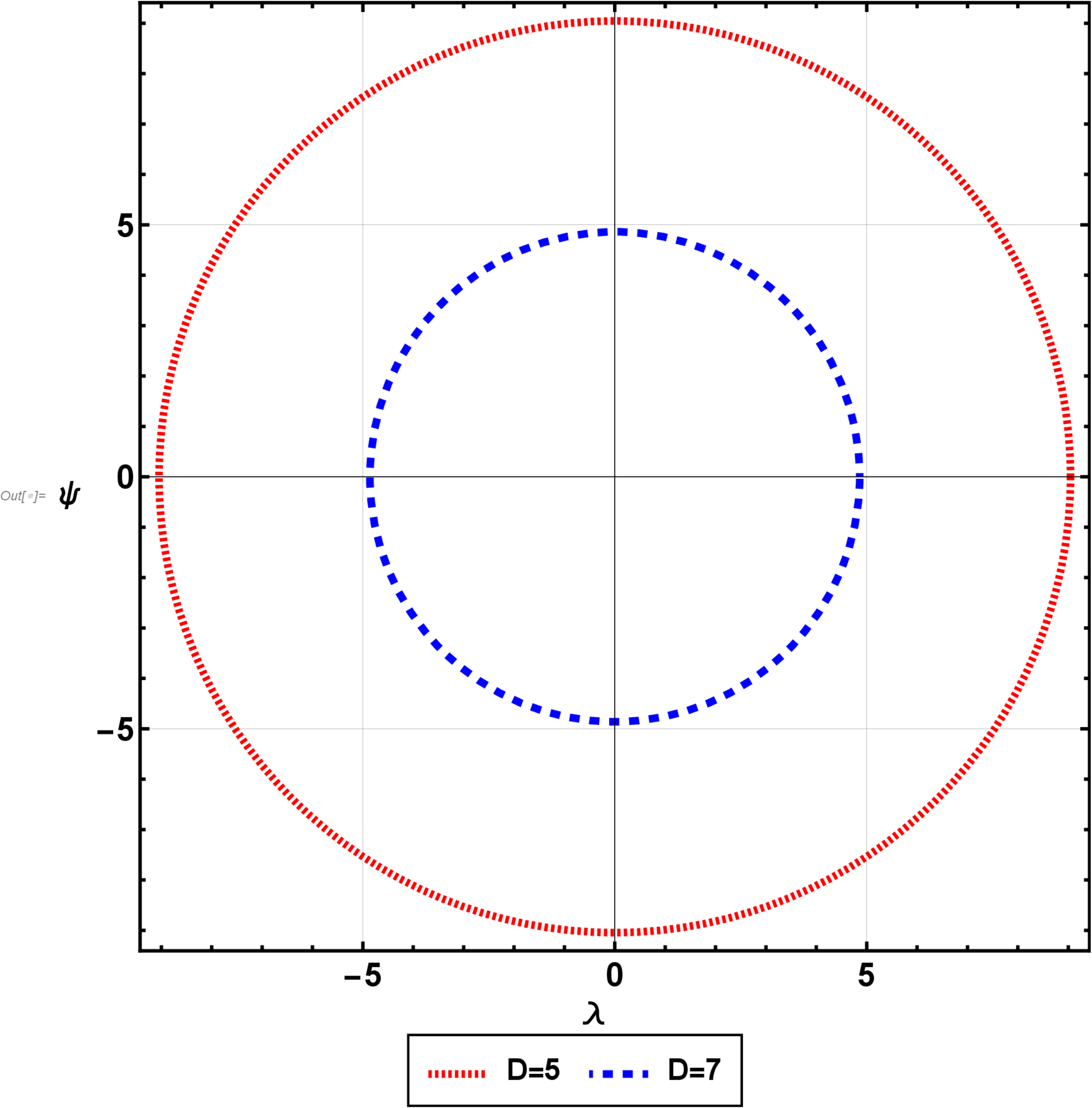}
 \caption{\small
Shadow of the higher-dimensional loop quantum de Sitter black hole in celestial coordinates for different spacetime dimensions $D$. The parameters are fixed at $\gamma=0.18$, and $\Lambda=0.10$.
} \label{figure5}
\end{figure}

\begin{figure}[p]
\centering

\subfloat[$D=4$]{%
\includegraphics[width=0.35\textwidth]{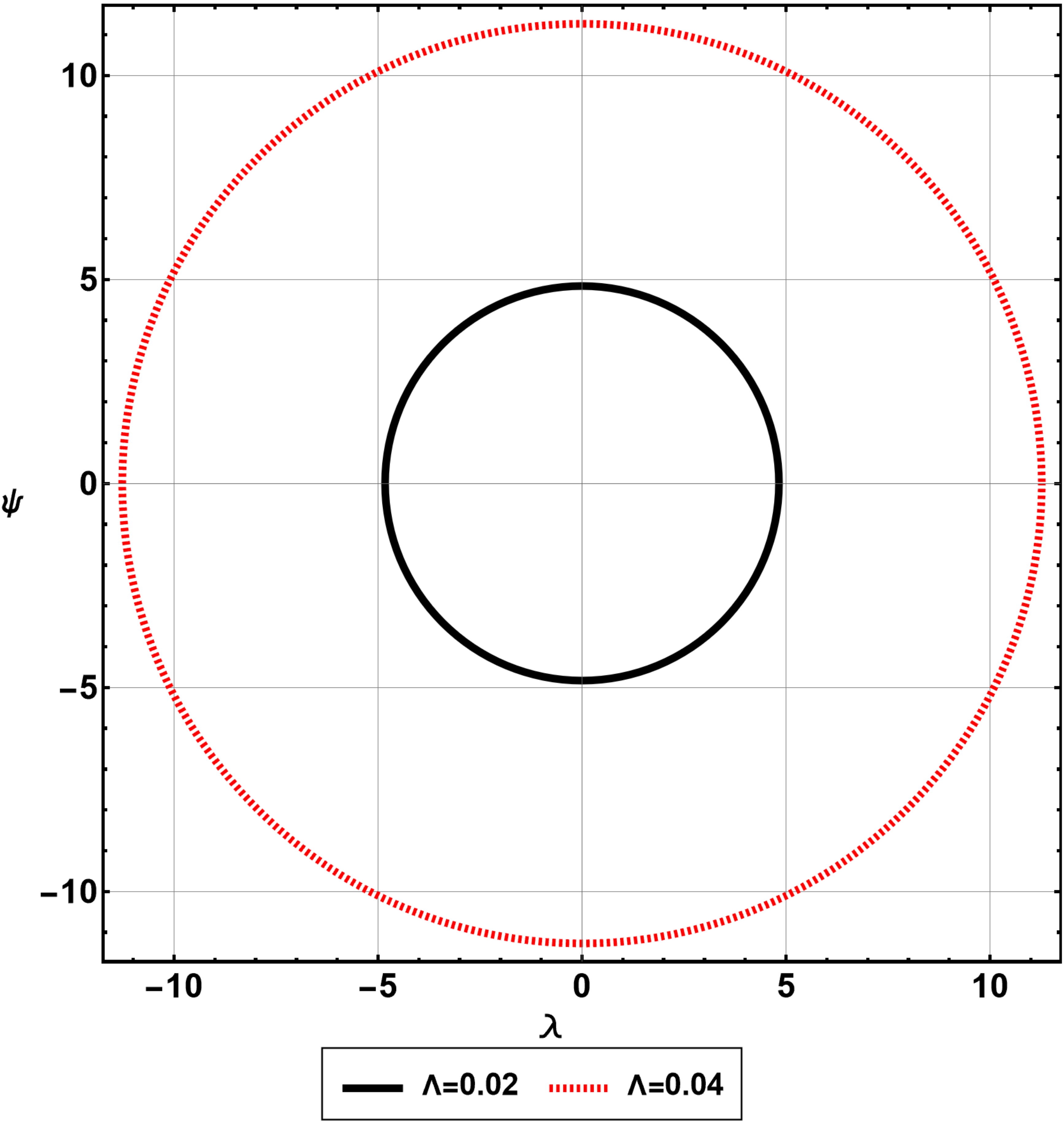}
}\\[-0.05cm]

\subfloat[$D=5$]{%
\includegraphics[width=0.35\textwidth]{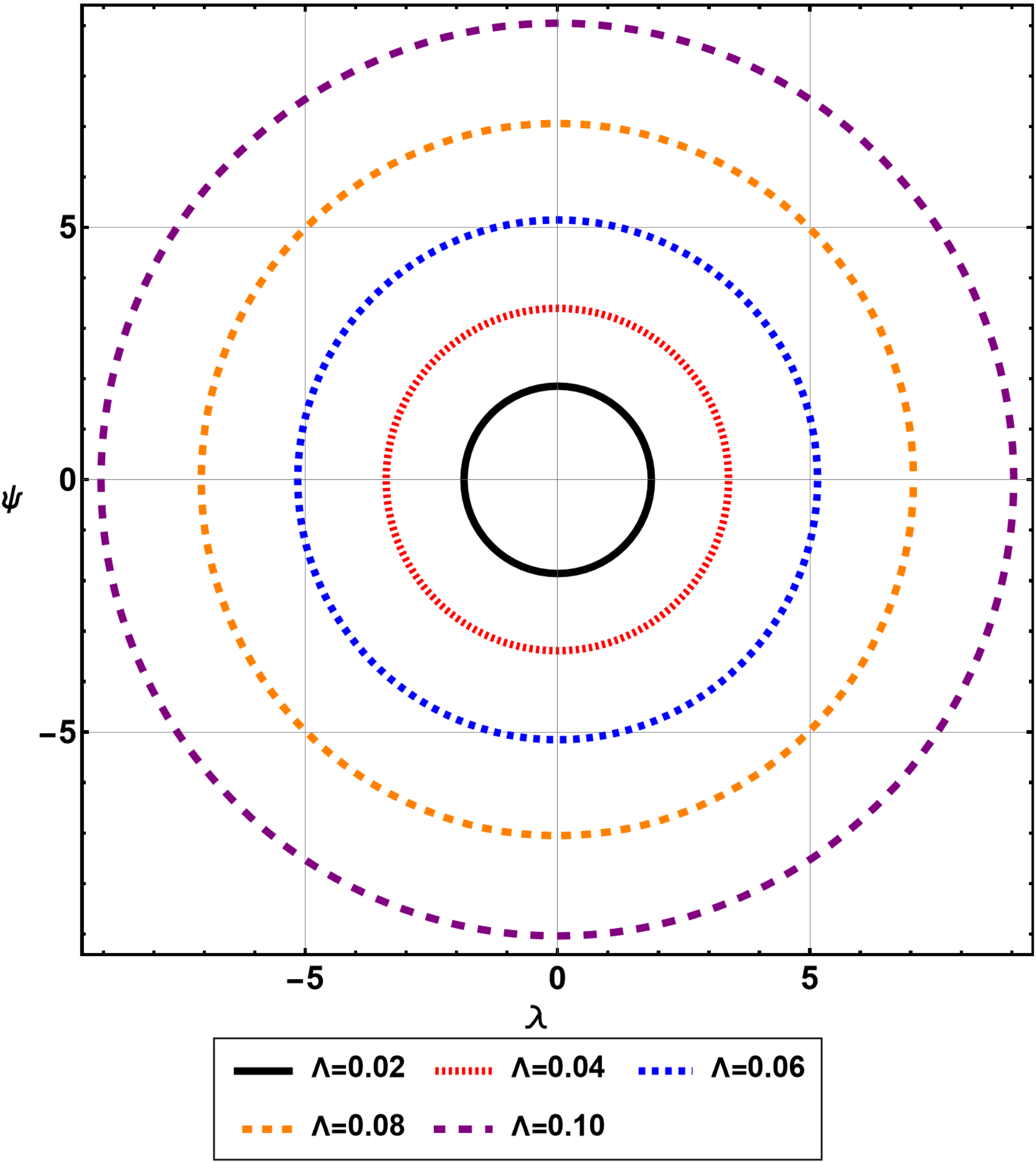}
}\\[-0.05cm]

\subfloat[$D=7$]{%
\includegraphics[width=0.35\textwidth]{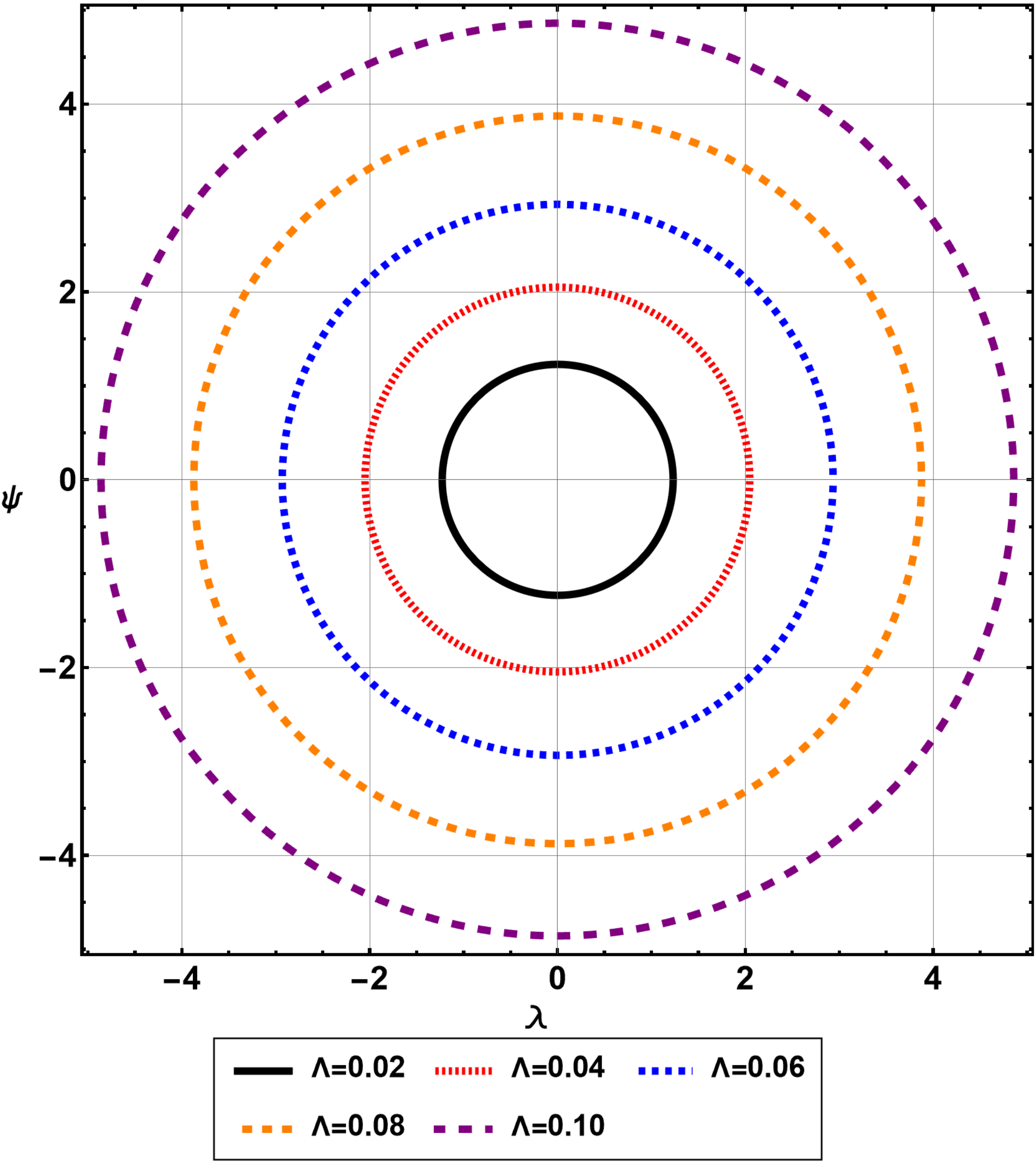}
}

\caption{\small
Shadow profiles of the higher-dimensional loop quantum de Sitter black hole for different values of the cosmological constant $\Lambda$ and spacetime dimensions $D$. The loop quantum parameter is fixed at $\gamma=0.18$.
}
\label{figure6}
\end{figure}

The qualitative features of the shadow are illustrated in Figs.~\ref{figure4}, \ref{figure5} and \ref{figure6}. These figures provide a direct visualization of how the underlying spacetime parameters affect the optical appearance of the black hole.\\

Figure~\ref{figure4} displays the shadow of the four-dimensional loop quantum de Sitter black hole for different values of the loop quantum parameter $\gamma$. The shadow remains perfectly circular due to the spherical symmetry of the spacetime, while its radius exhibits a clear dependence on the strength of the quantum correction. As $\gamma$ increases, the shadow boundary progressively contracts, indicating a reduction in the critical impact parameter that separates captured and escaping photon trajectories. This behavior is consistent with the results reported in Table~III, where both the photon-sphere radius $r_{ph}$ and the shadow radius $R_{\rm Sh}$ decrease monotonically with increasing $\gamma$. Physically, stronger loop quantum corrections modify the near-horizon geometry and shift the unstable photon orbit toward smaller radii, thereby reducing the extent of the photon capture region. Consequently, photons can approach the black hole more closely before being trapped, leading to a smaller apparent shadow. The effect becomes particularly pronounced for larger values of $\gamma$, where the contraction of the shadow is substantial compared to the weakly corrected regime.

The influence of extra spatial dimensions is illustrated in Fig.~\ref{figure5}, where shadow profiles are shown for different spacetime dimensions while keeping $\gamma=0.18$ and $\Lambda=0.10$ fixed. A pronounced reduction of the shadow size is observed as the dimensionality increases. In particular, the shadow corresponding to $D=7$ is considerably smaller than that for $D=5$, demonstrating that the dimensional dependence is significantly stronger than the effect produced by moderate variations of the loop parameter. This behavior originates from the modification of the gravitational field in higher-dimensional spacetimes, which shifts the unstable circular photon orbit inward and decreases the associated critical impact parameter. As a result, the region of phase space occupied by captured photon trajectories shrinks, producing a substantially smaller shadow. Despite this strong quantitative change in size, the shadow preserves its circular shape, indicating that higher dimensions alter the scale of the photon sphere rather than introducing any angular distortion. The trend is fully consistent with the values reported in Table~IV, where both the photon-sphere radius and the shadow radius decrease with increasing spacetime dimension.

The combined effects of the cosmological constant and spacetime dimensionality are presented in Fig.~\ref{figure6}. For each dimension, increasing the cosmological constant leads to a systematic enlargement of the shadow radius. This trend reflects the outward displacement of the unstable photon orbit induced by the de Sitter background, which increases the critical impact parameter and consequently enlarges the apparent shadow. The effect is particularly significant in higher dimensions, where the shadow radius grows rapidly as $\Lambda$ increases from $0.02$ to $0.10$. Nevertheless, for any fixed value of $\Lambda$, higher-dimensional configurations always produce smaller shadows than their lower-dimensional counterparts. Figure~\ref{figure5} therefore reveals a clear competition between two effects: the cosmological constant tends to enlarge the shadow, whereas extra dimensions act to suppress it. The observed shadow size is ultimately determined by the interplay between these two mechanisms together with the loop quantum corrections encoded in $\gamma$.

These results are fully consistent with the behavior observed in the quasinormal mode spectrum, where both the oscillation frequency and damping rate increase with dimension. Since both phenomena are governed by the same photon sphere structure, this agreement provides a strong consistency check of our analysis.\\

\noindent\textbf{Comparison with the classical shadow.}\quad
The shadow results may also be interpreted relative to the classical
Schwarzschild limit. For a four-dimensional Schwarzschild black hole,
the unstable circular photon orbit is located at
$r_{\rm ph}=3M$ and the corresponding critical impact parameter is
$R_{\rm Sh}=3\sqrt{3}\,M$. The present quantum-corrected geometry
approaches the corresponding classical behavior as the quantum
deformation is reduced, while finite values of $\gamma$ modify the
photon-sphere radius and hence the critical impact parameter defining
the shadow. As shown by the numerical results above, increasing
$\gamma$ shifts the unstable photon orbit inward and decreases
$R_{\rm Sh}$. Thus, in four dimensions and for sufficiently weak
quantum corrections, the shadow remains close to the classical
black-hole prediction, while progressively larger quantum corrections
produce an increasing departure from it.

For the de Sitter configurations considered in this work, however,
the comparison with Schwarzschild should be interpreted with some
care, since the cosmological constant itself also modifies the optical
geometry. At fixed $\Lambda$, the Schwarzschild--de Sitter solution
therefore provides the more direct classical reference for isolating
the loop-quantum contribution, whereas Schwarzschild represents the
asymptotically flat GR benchmark. In higher dimensions the analogous
classical reference is the Tangherlini--de Sitter geometry. The
numerical results indicate that the change associated with spacetime
dimensionality can be substantially larger than the moderate
loop-quantum correction. This separation of effects is particularly
relevant when confronting the model with shadow observations, since
small departures from the classical prediction may remain within
current observational uncertainties, whereas sufficiently large
changes of the photon-sphere scale can be constrained by the EHT data.

\subsection{Constraints from EHT observations of $M87^{\ast}$} \label{se5}

The latest observations by the Event Horizon Telescope (EHT) of the supermassive black hole $\mathrm{M87}^*$ have provided a new scientific opportunity to test theories of gravity through observable quantities associated with the black hole shadow, significantly advancing our understanding of physics in strong gravitational fields. In recent years, published articles and research resources have modeled and constrained black hole shadows using various black hole candidates and observational data. In this section, we model $\mathrm{M87}^*$ as a quantum-corrected higher dimensional black hole and apply the bounds obtained by EHT on the shadow diameter $d_{\mathrm{sh}}$ to extract constraints on the model parameters.

The EHT collaboration has reported that for the supermassive black hole $\mathrm{M87}^*$, the angular diameter of the shadow $\theta_{\mathrm{M87}^*}$, the mass $M_{\mathrm{M87}^*}$, and the distance $D_{\mathrm{M87}^*}$ from Earth are given by \cite{akiyama2019first}:
\begin{equation}
\theta_{\mathrm{M87}^*} = (42 \pm 3)\mu\mathrm{as}, \qquad 
M_{\mathrm{M87}^*} = (6.5 \pm 0.9) \times 10^{9} M_{\odot}, \qquad 
\mathcal{D}_{\mathrm{M87}^*} = 16.8^{+0.8}_{-0.7} \mathrm{Mpc},
\label{eq:EHT_params}
\end{equation}
where $M_{\odot}$ denotes the solar mass. Using these reported values, the shadow diameter in units of mass is constrained as \cite{akiyama2019first}:
\begin{equation}
d_{\mathrm{M87}^*} \equiv \frac{\mathcal{D}\theta}{M} \approx 11.0 \pm 1.5.
\label{eq:shadow_diameter}
\end{equation}
In this relation, $\mathcal{D}$ represents the distance from Earth, $\theta$ is the angular diameter of the shadow, and $M$ is the black hole mass.

As can be seen from Eq.~\eqref{eq:shadow_diameter}, at the $1\sigma$ confidence level we have
\begin{equation}
\label{dm87}
9.5 \lesssim d_{\mathrm{M87}^*} \lesssim 12.5,
\end{equation}
while at the $2\sigma$ confidence level, this quantity lies within the range
\begin{equation}
8 \lesssim d_{\mathrm{M87}^*} \lesssim 14.
\end{equation}
We now compare the shadow radius of the higher-dimensional de Sitter loop quantum black hole with the observed shadow size of the supermassive black hole M87$^{\ast}$ measured by the Event Horizon Telescope (EHT), as given in Eq.~\eqref{dm87}. This comparison allows us to place constraints on the cosmological constant $\Lambda$ and the loop quantum parameter $\gamma$.\\

\begin{figure}[p]
\centering

\subfloat[$\gamma=0.18$]{%
\includegraphics[width=0.65\textwidth]{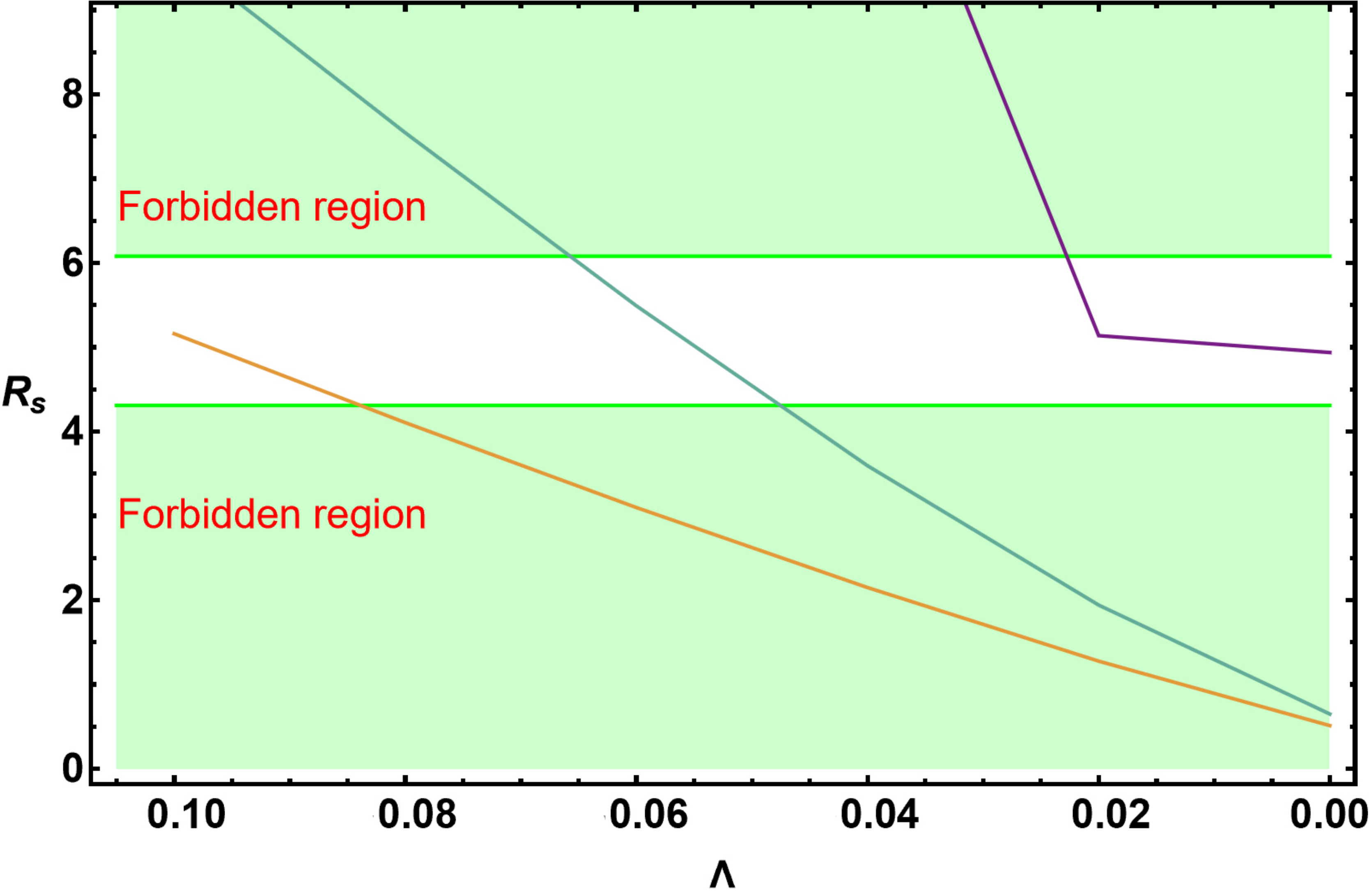}
\label{a}
}\\[0.4cm]

\subfloat[$\gamma=0.21$]{%
\includegraphics[width=0.65\textwidth]{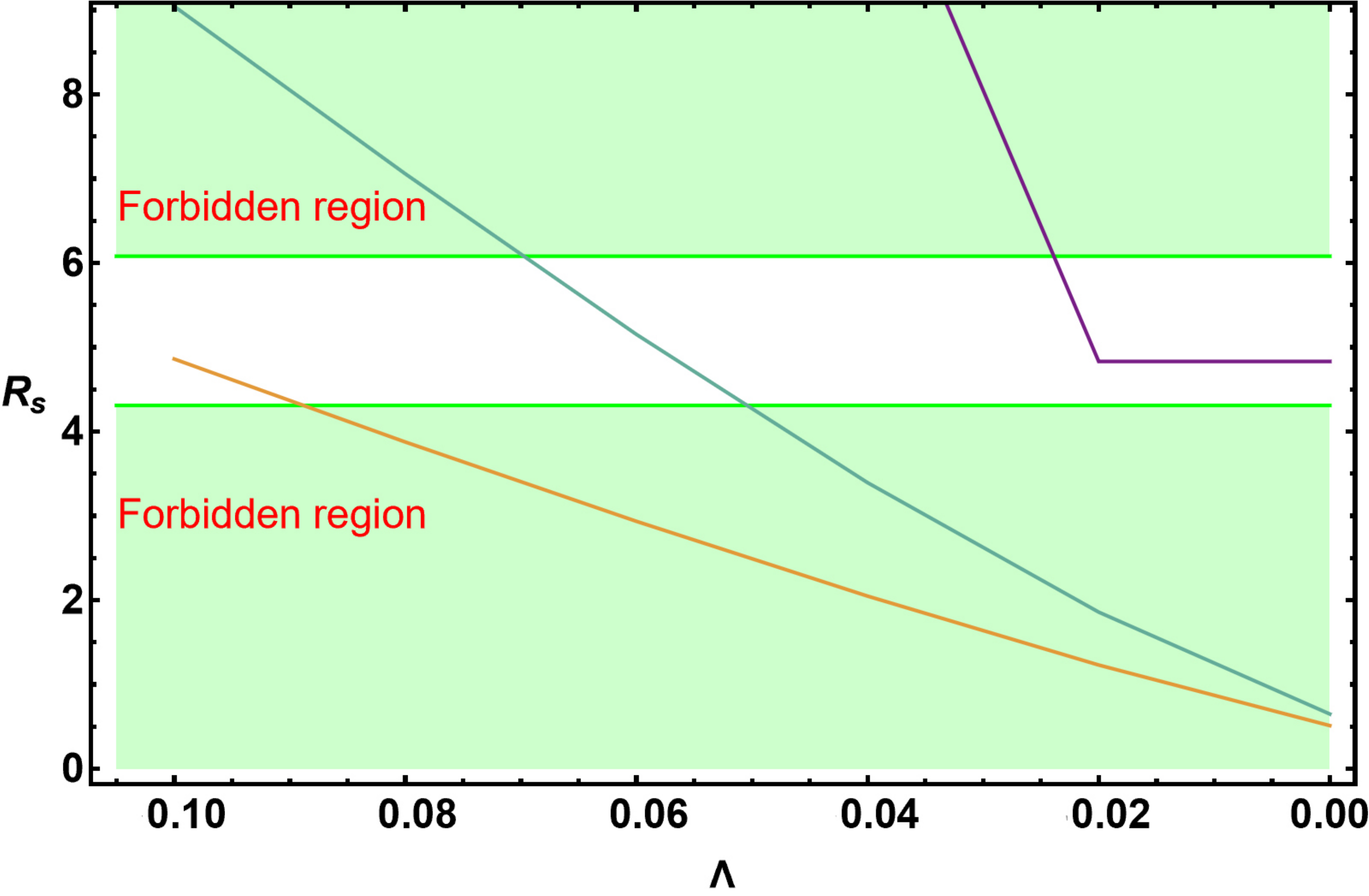}
\label{b}
}\\[0.4cm]

\subfloat[$\Lambda=0.1$]{%
\includegraphics[width=0.65\textwidth]{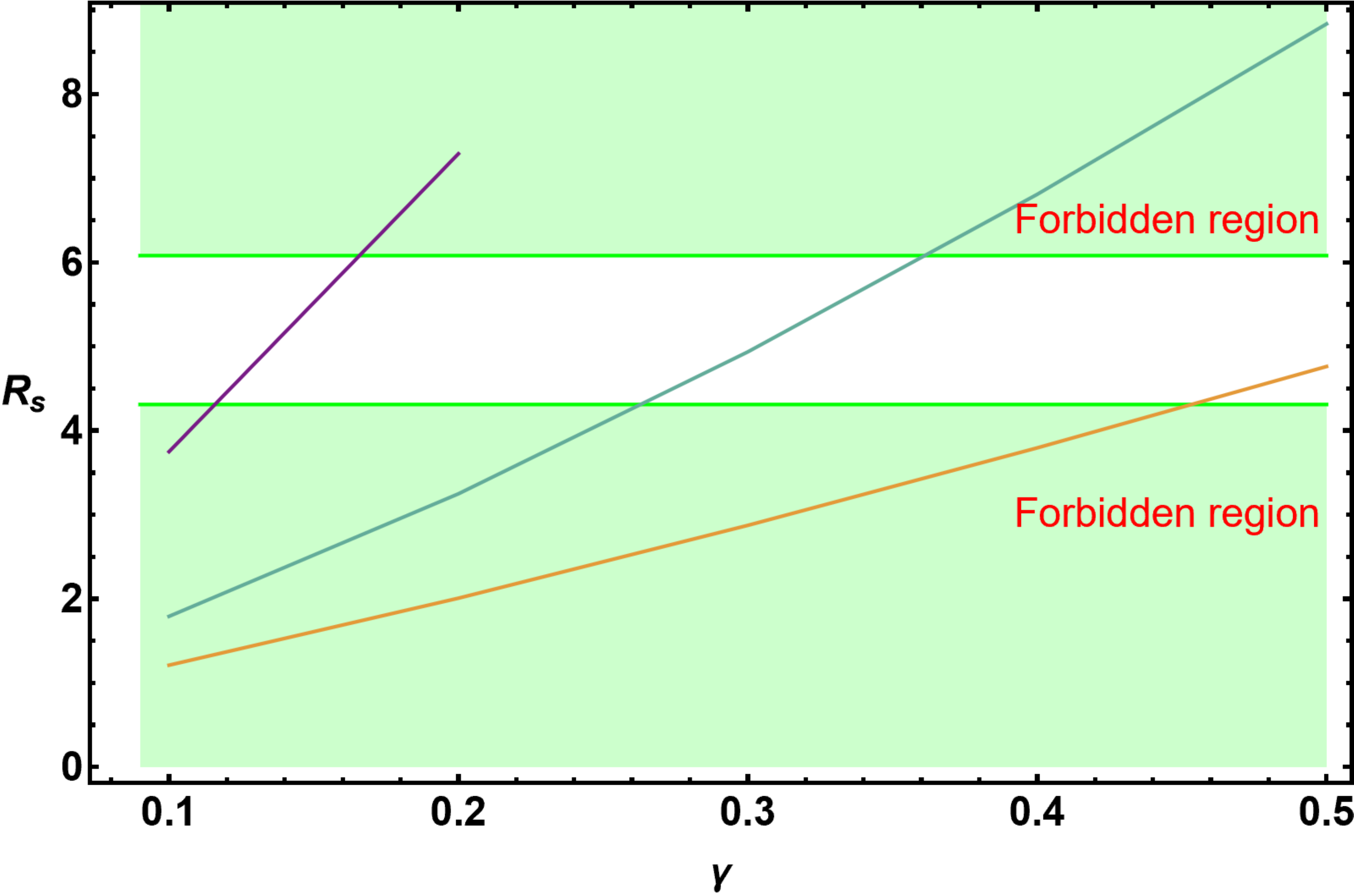}
\label{c}
}

\caption{\small Constraints on the parameter space of the loop quantum de Sitter black hole from EHT observations of $M87^{\ast}$. The shaded regions are excluded, while the unshaded regions correspond to parameter values consistent with observational bounds. The purple, green, and orange curves correspond to $D=4$, $D=5$, and $D=7$, respectively.}
\label{fig:quantum_shadow}
\end{figure}

Figure~\ref{fig:quantum_shadow} illustrates the behavior of the shadow radius of the higher-dimensional de Sitter loop quantum black hole in comparison with the observational bounds of M87$^{\ast}$ within the $1$-$\sigma$ uncertainty range. In this figure, the white (unshaded) region represents the allowed parameter space consistent with EHT observations, while the green (shaded) regions correspond to excluded configurations that are incompatible with the measured shadow size. In Figs.~\ref{a} and \ref{b}, we present the dependence of the shadow radius on the cosmological constant $\Lambda$ for two representative values of the loop parameter, namely $\gamma=0.21$ and $\gamma=0.18$. For $\gamma=0.21$, Fig.~\ref{a} shows that for $D=4$ the shadow radius is consistent with the M87$^{\ast}$ observations within the range $0<\Lambda<0.025$, while for $D=5$ and $D=7$ the allowed regions shift to $0.05<\Lambda<0.07$ and $\Lambda>0.09$, respectively. Comparing Figs.~\ref{a} and \ref{b}, one observes that decreasing $\gamma$ slightly shifts the allowed ranges toward larger values of $\Lambda$, while increasing $\gamma$ leads to a mild decrease of the shadow radius. These figures also indicate that neglecting the cosmological constant results in a smaller shadow size.
Figure~\ref{c} shows the dependence of the shadow radius on the loop parameter $\gamma$ for a fixed value of the cosmological constant, $\Lambda=0.10$. In this case, the compatibility with M87$^{\ast}$ observations occurs within the range $0.12<\gamma<0.17$ for $D=4$, while for $D=5$ and $D=7$ the allowed regions shift to $0.26<\gamma<0.36$ and $\gamma>0.46$, respectively. The figure clearly demonstrates that increasing the loop parameter $\gamma$ leads to an decrease in the shadow radius. An important implication of Fig.~\ref{fig:quantum_shadow} is that the presence of extra dimensions can leave observable imprints on the shadow of black holes. In particular, the case $D=5$ exhibits parameter ranges that are compatible with EHT observations, suggesting that higher-dimensional effects, in conjunction with the cosmological constant and loop quantum corrections, may be probed through black hole shadow measurements.

\section{summary and conclusion}
In this work, we have investigated the physical properties of a quantum-corrected black hole in a de Sitter background within the framework of loop quantum gravity, extending the construction to higher-dimensional spacetimes. This generalized model incorporates, in a unified manner, the effects of quantum geometry corrections, a cosmological constant, and extra spatial dimensions, providing a comprehensive setting for analyzing both dynamical and observational signatures of black holes.

Before addressing the perturbative and optical observables, we examined
the horizon structure of the quantum-corrected geometry and determined
the physical parameter domain in which it represents a nonextremal
de Sitter black hole. For fixed $(D,\gamma,\Lambda)$, the admissible
mass range is bounded from below by the extremal configuration, where
the inner and event horizons coincide, and from above by the Nariai
configuration, where the event and cosmological horizons merge. Since
these limiting masses depend on the spacetime dimension and the loop
parameter, the admissible mass range must be determined for each set of
model parameters. Throughout the subsequent analysis, the black-hole
mass was therefore restricted to the three-horizon sector
$r_{-}<r_{h}<r_{c}$, ensuring that the perturbative and optical
calculations are performed on physically admissible black-hole
backgrounds.

We next analyzed the response of the spacetime to massless scalar
perturbations. The corresponding effective potential was examined
within the physical static region $r_h<r<r_c$ and compared with the
Schwarzschild--de Sitter and Schwarzschild backgrounds in four
dimensions and with the Tangherlini--de Sitter background in five
dimensions. The quantum-corrected potentials retain the characteristic
single-barrier structure of their classical counterparts, while the
loop-quantum contribution modifies the height, position, and curvature
of the barrier, predominantly in the strong-field region. This provides
a direct geometrical interpretation of the quantitative shifts found in
the quasinormal spectrum. We computed the QNM spectrum using three independent approaches: time-domain evolution with multi-mode Prony extraction, the matrix (determinant) method, and the WKB approximation. The results obtained from these methods were found to be in excellent agreement, confirming the robustness and consistency of the numerical analysis. This consistency is also demonstrated directly in the time domain:
the numerical waveform exhibits a clear damped ringdown, and within
the optimized ringdown interval its oscillation phase and decay are
closely reproduced by both the Prony fit and the reconstruction based
on the independently obtained matrix-method frequency. Our findings show that loop quantum corrections introduce controlled quantitative modifications to both the oscillation frequency and damping rate of the modes, while preserving the overall structure of the photon-sphere-dominated spectrum. Moreover, we observed that the dependence of the quasinormal frequencies on the spacetime dimensions is significantly stronger than on the loop parameter, with both the real and imaginary parts increasing rapidly as the number of spatial dimensions grows. This behavior can be attributed to the inward shift of the photon sphere and the steepening of the effective potential barrier in higher dimensions. The comparison with the corresponding classical backgrounds further
shows that the loop-quantum-corrected modes remain continuously
connected to the usual photon-sphere quasinormal branch. Thus, within
the parameter range investigated here, the quantum correction acts
mainly as a controlled deformation of the classical ringdown spectrum,
whereas the change in spacetime dimensionality produces the dominant
shift in both the oscillation and damping scales. An important result of our analysis is that the imaginary parts of all quasinormal frequencies remain negative throughout the parameter space investigated, indicating the absence of exponentially growing modes. Therefore, the higher-dimensional loop quantum de Sitter black holes considered in this work are dynamically stable against massless scalar perturbations within the explored range of parameters.

In addition to the dynamical analysis, we explored the optical properties of the black hole by studying null geodesics and constructing the corresponding shadow. We showed that the shadow radius is strongly influenced by the interplay between the loop quantum parameter, the cosmological constant, and the number of spacetime dimensions. In particular, increasing the loop parameter generally leads to a reduction in the shadow size, while a positive cosmological constant tends to enlarge it. The effect of extra dimensions is especially pronounced, as higher-dimensional spacetimes exhibit a substantial decrease in the shadow radius due to the modified structure of photon trajectories. Relative to the classical Schwarzschild reference, these modifications
may be understood as controlled deformations of the photon-sphere
geometry rather than a qualitatively different optical structure.
Accordingly, the Schwarzschild limit provides a useful benchmark for
quantifying the magnitude of the loop-quantum and higher-dimensional
corrections to the shadow.

A key aspect of this work is the comparison of the theoretical shadow radius with observational constraints from the Event Horizon Telescope measurements of the supermassive black hole M87$^{\ast}$. By confronting our model with the observed shadow size within the reported uncertainty bounds, we were able to place constraints on the parameter space of the cosmological constant and the loop quantum parameter. Our analysis indicates that certain ranges of these parameters, particularly in the presence of extra spatial dimensions, remain consistent with observational data. Notably, the results suggest that higher-dimensional effects, especially for specific dimensionalities, may leave observable imprints on black hole shadows and could, in principle, be probed by current or future high-resolution observations.

Overall, our study demonstrates that the combined effects of loop quantum gravity corrections, a cosmological constant, and extra dimensions lead to distinctive and physically consistent modifications in both the QNM spectrum and the shadow characteristics of black holes. The close agreement between different computational methods, together with the compatibility of the model with observational data, provides strong support for the reliability of the results. These findings highlight the potential of black hole spectroscopy and shadow observations as powerful tools for probing beyond-standard physics, including quantum gravitational effects and the possible existence of extra spatial dimensions.

The dynamical stability demonstrated by the QNM analysis, together with the compatibility of the shadow observables with EHT measurements, supports the physical viability of the higher-dimensional loop quantum de Sitter black hole model investigated in this work. Future studies may extend the present analysis to rotating quantum-corrected black hole solutions, examine other classes of perturbations such as electromagnetic and gravitational modes, and explore more general cosmological backgrounds. Furthermore, the incorporation of increasingly precise observational data from next-generation very-long-baseline interferometry and gravitational-wave experiments could provide significantly tighter constraints on the model parameters, thereby improving the prospects for identifying observational signatures of loop quantum gravity effects and extra spatial dimensions in astrophysical black holes.

\begin{acknowledgments}

The authors gratefully acknowledge Jerzy Lewandowski,  Yongge Ma, Jinsong Yang and Cong Zhang for kindly providing their detailed calculations related to the derivation of the metric via the junction conditions. Their helpful explanations and generous correspondence significantly assisted us in verifying and completing the derivation presented in this work. The Work of KN is supported financially by the INSF of Iran under the grant number 40501739, and SS is supported
financially by the INSF of Iran under the grant number 40408889
\end{acknowledgments}

\appendix

\section{Derivation of the exterior metric from the junction conditions
and its higher-dimensional extension}
\label{app:junction}

In this Appendix, we summarize the junction-condition construction
leading to the four-dimensional quantum-corrected exterior metric and
its extension to $D$ dimensions. The loop-quantum-cosmology ingredients
entering the interior dynamics, including the effective Friedmann
equation, the critical density, and the area-gap parameter, have already
been introduced in Sec.~II and will not be repeated here.

\subsection{Four-dimensional junction construction}

Following the quantum Oppenheimer--Snyder construction of
Ref.~\cite{lewandowski2023quantum}, the homogeneous dust interior is
described by
\begin{equation}
ds_{\rm in}^{2}
=
-d\tau^{2}
+
a^{2}(\tau)
\left(
d\tilde r^{\,2}
+
\tilde r^{\,2}d\Omega^{2}
\right),
\label{A1}
\end{equation}
with the dust surface located at the fixed comoving coordinate
$\tilde r=\tilde r_{0}$. Its physical radius is therefore
$a(\tau)\tilde r_{0}$.

The exterior geometry is assumed to be static and spherically
symmetric,
\begin{equation}
ds_{\rm out}^{2}
=
-[1-F(r)]\,dt^{2}
+
[1-G(r)]^{-1}dr^{2}
+
r^{2}d\Omega^{2},
\label{A2}
\end{equation}
where the functions $F(r)$ and $G(r)$ are determined by matching the
two geometries across the dust surface. The junction hypersurface is
identified according to
\begin{equation}
(\tau,\tilde r_{0},\theta,\phi)
\sim
(t(\tau),r(\tau),\theta,\phi).
\label{A3}
\end{equation}

The trajectory of the boundary in the exterior geometry is a radial
timelike geodesic. Conservation of its norm and of the Killing energy
gives
\begin{equation}
-1
=
-[1-F(r)]\dot t^{\,2}
+
[1-G(r)]^{-1}\dot r^{\,2},
\label{A4}
\end{equation}
and
\begin{equation}
E=[1-F(r)]\dot t .
\label{A5}
\end{equation}
Consequently,
\begin{equation}
\dot t
=
\frac{E}{1-F(r)},
\qquad
\dot r^{\,2}
=
E^{2}\frac{1-G(r)}{1-F(r)}
-1+G(r).
\label{A6}
\end{equation}

The Darmois--Israel matching conditions require continuity of the
induced metric and extrinsic curvature across the junction surface,
\begin{equation}
[h_{\mu\nu}]=0,
\qquad
[K_{\mu\nu}]=0.
\label{A7}
\end{equation}
They lead to
\begin{equation}
r(\tau)=a(\tau)\tilde r_{0},
\label{A8}
\end{equation}
together with
\begin{equation}
E^{2}
\frac{1-G(r)}{1-F(r)}
=1.
\label{A9}
\end{equation}
Using Eq.~(\ref{A9}) in the radial geodesic equation yields
\begin{equation}
\dot r^{\,2}
=
\dot a^{\,2}\tilde r_{0}^{\,2}
=
G(r).
\label{A10}
\end{equation}

Substituting the effective four-dimensional Friedmann equation given
in Sec.~II and using $r=a(\tau)\tilde r_{0}$, one obtains
\begin{equation}
G(r)
=
\frac{2GM}{r}
-
\frac{\alpha G^{2}M^{2}}{r^{4}}
\left(
1+\frac{\Lambda r^{3}}{6GM}
\right)^{2}
+
\frac{\Lambda r^{2}}{3}.
\label{A11}
\end{equation}

The radial coordinate followed by the dust surface satisfies
$r\in[r_{b},\infty)$, where $r_{b}$ is the minimum radius reached at
the bounce. Equation~(\ref{A9}) further implies that, after the constant
rescaling $t\rightarrow t/E$, the exterior line element may be written
as
\begin{equation}
ds_{\rm out}^{2}
=
-f(r)\,dt^{2}
+
\frac{dr^{2}}{f(r)}
+
r^{2}d\Omega^{2},
\label{A12}
\end{equation}
with
\begin{equation}
f(r)
=
1
-
\frac{2GM}{r}
-
\frac{\Lambda r^{2}}{3}
+
\frac{\alpha G^{2}M^{2}}{r^{4}}
\left(
1+\frac{\Lambda r^{3}}{6GM}
\right)^{2}.
\label{A13}
\end{equation}
This reproduces the four-dimensional metric employed in Sec.~II.

\subsection{Extension to $D$ dimensions}

The same construction can be extended to a homogeneous and isotropic
dust interior in $D$ spacetime dimensions,
\begin{equation}
ds_{\rm in}^{2}
=
-d\tau^{2}
+
a^{2}(\tau)
\left(
d\tilde r^{\,2}
+
\tilde r^{\,2}d\Omega_{D-2}^{2}
\right).
\label{A14}
\end{equation}
The area of the unit $(D-2)$-sphere is
\begin{equation}
\omega
=
\frac{2\pi^{(D-1)/2}}
{\Gamma\!\left[(D-1)/2\right]},
\label{A15}
\end{equation}
and it is convenient to introduce
\begin{equation}
m
=
\frac{8\pi M}
{(D-2)\omega}.
\label{A16}
\end{equation}

As discussed in Sec.~II, the effective higher-dimensional LQC
dynamics is governed by
\begin{equation}
H^{2}
=
\frac{16\pi G}
{(D-1)(D-2)}
\rho
\left(
1-\frac{\rho}{\rho_{c}}
\right),
\label{A17}
\end{equation}
where
\begin{equation}
\rho_{c}
=
\frac{(D-1)(D-2)}
{16\pi G\gamma^{2}
\Delta^{2/(D-2)}}.
\label{A18}
\end{equation}
Here $\gamma$ is the Barbero--Immirzi parameter and
\begin{equation}
\Delta
=
8\sqrt{D-1}\,\pi\gamma\,
(l_{p})^{D-1},
\qquad
l_{p}
=
\sqrt{(D-2)G\hbar},
\label{A19}
\end{equation}
in the convention adopted in Sec.~II.

Applying the same junction procedure as in
Eqs.~(\ref{A3})--(\ref{A10}) to the $D$-dimensional interior gives a
static and spherically symmetric exterior,
\begin{equation}
ds_{\rm out}^{2}
=
-h(r)\,dt^{2}
+
\frac{dr^{2}}{f(r)}
+
r^{2}d\Omega_{D-2}^{2},
\label{A20}
\end{equation}
with
\begin{equation}
\begin{aligned}
h(r)=f(r)
={}&
1
-
\left[
\frac{2m}{r^{D-3}}
-
\frac{
4\gamma^{2}
\Delta^{2/(D-2)}
m^{2}}
{r^{4(D-3)}}
\right]
\\
&\times
\left[
1+
\frac{2\Lambda r^{3}}
{m(D-1)(D-2)}
\right]^{2}
-
\frac{2\Lambda r^{2}}
{(D-1)(D-2)} .
\end{aligned}
\label{A21}
\end{equation}

Equation~(\ref{A21}) is the higher-dimensional
loop-quantum-corrected de Sitter metric used throughout the main text.
For $D=4$, it consistently reduces to the four-dimensional form in
Eq.~(\ref{A13}).

\bibliographystyle{apsrev4-2}
\bibliography{Reff}

@article{rovelli1995discreteness,
  title={Discreteness of area and volume in quantum gravity},
  author={Rovelli, Carlo and Smolin, Lee},
  journal={Nuclear Physics B},
  volume={442},
  number={3},
  pages={593--619},
  year={1995},
  publisher={Elsevier}
}

@article{ashtekar1997quantum,
  title={Quantum theory of geometry II: Volume operators},
  author={Ashtekar, Abhay and Lewandowski, Jerzy},
  journal={arXiv preprint gr-qc/9711031},
  year={1997}
}

@article{han2007fundamental,
  title={Fundamental structure of loop quantum gravity},
  author={Han, Muxin and Ma, Yongge and Huang, Weiming},
  journal={International Journal of Modern Physics D},
  volume={16},
  number={09},
  pages={1397--1474},
  year={2007},
  publisher={World Scientific}
}

@article{ashtekar2011loop,
  title={Loop quantum cosmology: a status report},
  author={Ashtekar, Abhay and Singh, Parampreet},
  journal={Classical and Quantum Gravity},
  volume={28},
  number={21},
  pages={213001},
  year={2011}
}

@article{zhang2022first,
  title={First-order quantum correction in coherent state expectation value of loop-quantum-gravity Hamiltonian},
  author={Zhang, Cong and Song, Shicong and Han, Muxin},
  journal={Physical Review D},
  volume={105},
  number={6},
  pages={064008},
  year={2022},
  publisher={APS}
}

@article{zhang2023fermions,
  title={Fermions in loop quantum gravity and resolution of doubling problem},
  author={Zhang, Cong and Liu, Hongguang and Han, Muxin},
  journal={Classical and Quantum Gravity},
  volume={40},
  number={20},
  pages={205022},
  year={2023},
  publisher={IOP Publishing}
}

@article{chiou2008phenomenological,
  title={Phenomenological loop quantum geometry of the Schwarzschild black hole},
  author={Chiou, Dah-Wei},
  journal={Physical Review D—Particles, Fields, Gravitation, and Cosmology},
  volume={78},
  number={6},
  pages={064040},
  year={2008},
  publisher={APS}
}

@article{gambini2008black,
  title={Black holes in loop quantum gravity: the complete space-time},
  author={Gambini, Rodolfo and Pullin, Jorge},
  journal={Physical review letters},
  volume={101},
  number={16},
  pages={161301},
  year={2008},
  publisher={APS}
}

@article{haggard2015quantum,
  title={Quantum-gravity effects outside the horizon spark black to white hole tunneling},
  author={Haggard, Hal M and Rovelli, Carlo},
  journal={Physical Review D},
  volume={92},
  number={10},
  pages={104020},
  year={2015},
  publisher={APS}
}

@article{christodoulou2016realistic,
  title={Realistic observable in background-free quantum gravity: the Planck-star tunnelling-time},
  author={Christodoulou, Marios and Rovelli, Carlo and Speziale, Simone and Vilensky, Ilya},
  journal={arXiv preprint arXiv:1605.05268},
  year={2016}
}

@article{ashtekar2018quantum,
  title={Quantum transfiguration of Kruskal black holes},
  author={Ashtekar, Abhay and Olmedo, Javier and Singh, Parampreet},
  journal={Physical review letters},
  volume={121},
  number={24},
  pages={241301},
  year={2018},
  publisher={APS}
}

@article{zhang2020loop,
  title={Loop quantum Schwarzschild interior and black hole remnant},
  author={Zhang, Cong and Ma, Yongge and Song, Shupeng and Zhang, Xiangdong},
  journal={Physical Review D},
  volume={102},
  number={4},
  pages={041502},
  year={2020},
  publisher={APS}
}

@article{zhang2022loop,
  title={Loop quantum deparametrized Schwarzschild interior and discrete black hole mass},
  author={Zhang, Cong and Ma, Yongge and Song, Shupeng and Zhang, Xiangdong},
  journal={Physical Review D},
  volume={105},
  number={2},
  pages={024069},
  year={2022},
  publisher={APS}
}

@article{lewandowski2023quantum,
  title={Quantum oppenheimer-snyder and swiss cheese models},
  author={Lewandowski, Jerzy and Ma, Yongge and Yang, Jinsong and Zhang, Cong},
  journal={Physical Review Letters},
  volume={130},
  number={10},
  pages={101501},
  year={2023},
  publisher={APS}
}

@article{husain2022fate,
  title={Fate of quantum black holes},
  author={Husain, Viqar and Kelly, Jarod George and Santacruz, Robert and Wilson-Ewing, Edward},
  journal={Physical Review D},
  volume={106},
  number={2},
  pages={024014},
  year={2022},
  publisher={APS}
}

@article{stachowiak2007exact,
  title={Exact solutions in bouncing cosmology},
  author={Stachowiak, Tomasz and Szyd{\l}owski, Marek},
  journal={Physics Letters B},
  volume={646},
  number={5-6},
  pages={209--214},
  year={2007},
  publisher={Elsevier}
}

@article{ashtekar2006quantum2,
  title={Quantum geometry and the Schwarzschild singularity},
  author={Ashtekar, Abhay and Bojowald, Martin},
  journal={Classical and Quantum Gravity},
  volume={23},
  number={2},
  pages={391--411},
  year={2006}
}

@article{modesto2006loop,
  title={Loop quantum black hole},
  author={Modesto, Leonardo},
  journal={Classical and Quantum Gravity},
  volume={23},
  number={18},
  pages={5587--5601},
  year={2006}
}

@article{bojowald2018signature,
  title={Signature change in two-dimensional black-hole models of loop quantum gravity},
  author={Bojowald, Martin and Brahma, Suddhasattwa},
  journal={Physical Review D},
  volume={98},
  number={2},
  pages={026012},
  year={2018},
  publisher={APS}
}

@article{chiou2008phenomenological2,
  title={Phenomenological dynamics of loop quantum cosmology in Kantowski-Sachs spacetime},
  author={Chiou, Dah-Wei},
  journal={Physical Review D—Particles, Fields, Gravitation, and Cosmology},
  volume={78},
  number={4},
  pages={044019},
  year={2008},
  publisher={APS}
}

@article{zhang2023black,
  title={Black hole image encoding quantum gravity information},
  author={Zhang, Cong and Ma, Yongge and Yang, Jinsong},
  journal={Physical Review D},
  volume={108},
  number={10},
  pages={104004},
  year={2023},
  publisher={APS}
}

@article{yang2023shadow,
  title={Shadow and stability of quantum-corrected black holes},
  author={Yang, Jinsong and Zhang, Cong and Ma, Yongge},
  journal={The European Physical Journal C},
  volume={83},
  number={7},
  pages={619},
  year={2023},
  publisher={Springer}
}

@article{shao2024scalar,
  title={Scalar fields around a loop quantum gravity black hole in de Sitter spacetime: Quasinormal modes, late-time tails and strong cosmic censorship},
  author={Shao, Cai-Ying and Zhang, Cong and Zhang, Wei and Shao, Cheng-Gang},
  journal={Physical Review D},
  volume={109},
  number={6},
  pages={064012},
  year={2024},
  publisher={APS}
}

@article{regge1957stability,
  title={Stability of a Schwarzschild singularity},
  author={Regge, Tullio and Wheeler, John A},
  journal={Physical Review},
  volume={108},
  number={4},
  pages={1063},
  year={1957},
  publisher={APS}
}

@article{Horava:1995qa,
  author        = {Horava, Petr and Witten, Edward},
  title         = {{Heterotic and Type I string dynamics from eleven dimensions}},
  journal       = {Nucl. Phys. B},
  volume        = {460},
  pages         = {506--524},
  year          = {1996},
  doi           = {10.1016/0550-3213(95)00621-4},
  eprint        = {hep-th/9510209},
  archivePrefix = {arXiv}
}

@article{kaluza1921unitatsproblem,
  title={Zum unit{\"a}tsproblem der physik},
  author={Kaluza, Th},
  journal={Sitzungsber. Preuss. Akad. Wiss. Berlin (Math. Phys.)},
  volume={1921},
  number={arXiv: 1803.08616},
  pages={966--972},
  year={1921}
}

@article{klein1999quantum,
  title={Quantum theory and five-dimensional relativity theory},
  author={Klein, Oskar},
  journal={The Oskar Klein Memorial Lectures},
  volume={1},
  pages={67--80},
  year={1999},
  publisher={World Scientific}
}

@article{bailin1987kaluza,
  title={Kaluza-klein theories},
  author={Bailin, David and Love, Alex},
  journal={Reports on Progress in Physics},
  volume={50},
  number={9},
  pages={1087--1170},
  year={1987}
}

@article{witten1996five,
  title={Five-branes and M-theory on an orbifold},
  author={Witten, Edward},
  journal={Nuclear Physics B},
  volume={463},
  number={2-3},
  pages={383--397},
  year={1996},
  publisher={Elsevier}
}

@article{dvali20004d,
  title={4D gravity on a brane in 5D Minkowski space},
  author={Dvali, Gia and Gabadadze, Gregory and Porrati, Massimo},
  journal={Physics Letters B},
  volume={485},
  number={1-3},
  pages={208--214},
  year={2000},
  publisher={Elsevier}
}

@article{qiang2005five,
  title={Five-dimensional Brans-Dicke theory and cosmic acceleration},
  author={Qiang, Li-e and Ma, Yongge and Han, Muxin and Yu, Dan},
  journal={Physical Review D—Particles, Fields, Gravitation, and Cosmology},
  volume={71},
  number={6},
  pages={061501},
  year={2005},
  publisher={APS}
}

@article{zhang2013loop,
  title={Loop quantum Brans-Dicke cosmology},
  author={Zhang, Xiangdong and Artymowski, Michal and Ma, Yongge},
  journal={Physical Review D—Particles, Fields, Gravitation, and Cosmology},
  volume={87},
  number={8},
  pages={084024},
  year={2013},
  publisher={APS}
}

@article{leaver1986spectral,
  title={Spectral decomposition of the perturbation response of the Schwarzschild geometry},
  author={Leaver, Edward W},
  journal={Physical Review D},
  volume={34},
  number={2},
  pages={384},
  year={1986},
  publisher={APS}
}

@article{berti2009quasinormal,
  title={Quasinormal modes of black holes and black branes},
  author={Berti, Emanuele and Cardoso, Vitor and Starinets, Andrei O},
  journal={Classical and Quantum Gravity},
  volume={26},
  number={16},
  pages={163001},
  year={2009}
}

@article{abdalla2007perturbations,
  title={Perturbations of Schwarzschild black holes in laboratories},
  author={Abdalla, Elcio and Konoplya, RA and Zhidenko, A},
  journal={Classical and Quantum Gravity},
  volume={24},
  number={23},
  pages={5901--5909},
  year={2007}
}

@article{konoplya2007stability,
  title={Stability of multidimensional black holes: Complete numerical analysis},
  author={Konoplya, Roman A and Zhidenko, Alexander},
  journal={Nuclear Physics B},
  volume={777},
  number={1-2},
  pages={182--202},
  year={2007},
  publisher={Elsevier}
}

@article{zhidenko2008evolution,
  title={Evolution of Brane-Localised Standard Model Fields in Gauss-Bonnet theory},
  author={Zhidenko, Alexander},
  journal={arXiv preprint arXiv:0802.2262},
  year={2008}
}

@article{abdalla2007quasinormal,
  title={Quasinormal mode characterization of evaporating mini black holes},
  author={Abdalla, Elcio and Chirenti, Cecilia BMH and Saa, Alberto},
  journal={Journal of High Energy Physics},
  volume={2007},
  number={10},
  pages={086--086},
  year={2007}
}

@article{bizon2005critical,
  title={Critical behavior in vacuum gravitational collapse in 4+ 1 dimensions},
  author={Bizo{\'n}, Piotr and Chmaj, Tadeusz and Schmidt, Bernd G},
  journal={Physical review letters},
  volume={95},
  number={7},
  pages={071102},
  year={2005},
  publisher={APS}
}

@article{bizon2005vacuum,
  title={Vacuum gravitational collapse in nine dimensions},
  author={Bizo{\'n}, P and Chmaj, Tadeusz and Rostworowski, Andrzej and Schmidt, Bernd G and Tabor, Z},
  journal={Physical Review D—Particles, Fields, Gravitation, and Cosmology},
  volume={72},
  number={12},
  pages={121502},
  year={2005},
  publisher={APS}
}

@article{shi2024higher,
  title={Higher-dimensional quantum Oppenheimer-Snyder model},
  author={Shi, Zijian and Zhang, Xiangdong and Ma, Yongge},
  journal={Physical Review D},
  volume={110},
  number={10},
  pages={104074},
  year={2024},
  publisher={APS}
}

@article{kanti2004black,
  title={Black holes in theories with large extra dimensions: a review},
  author={Kanti, Panagiota},
  journal={International journal of modern physics A},
  volume={19},
  number={29},
  pages={4899--4951},
  year={2004},
  publisher={World Scientific}
}

@article{hod1998bohr,
  title={Bohr's correspondence principle and the area spectrum of quantum black holes},
  author={Hod, Shahar},
  journal={Physical Review Letters},
  volume={81},
  number={20},
  pages={4293},
  year={1998},
  publisher={APS}
}

@article{kunstatter2003d,
  title={d-dimensional black hole entropy spectrum from quasinormal modes},
  author={Kunstatter, Gabor},
  journal={Physical Review Letters},
  volume={90},
  number={16},
  pages={161301},
  year={2003},
  publisher={APS}
}

@article{panotopoulos2020quasinormal,
  title={Quasinormal modes of charged black holes in higher-dimensional Einstein-power-Maxwell theory},
  author={Panotopoulos, Grigoris},
  journal={Axioms},
  volume={9},
  number={1},
  pages={33},
  year={2020},
  publisher={MDPI}
}

@article{chabab2016behavior,
  title={Behavior of quasinormal modes and high dimension RN--AdS black hole phase transition},
  author={Chabab, M and Moumni, H El and Iraoui, S and Masmar, K},
  journal={The European Physical Journal C},
  volume={76},
  number={12},
  pages={676},
  year={2016},
  publisher={Springer}
}

@article{akiyamaetal2019eventhorizontelescope,
  title={EventHorizonTelescope},
  author={Akiyamaetal, K},
  journal={FirstM87eventhorizon telescope results. III. Data processing and calibration. Astrophys. J. Lett},
  volume={875},
  number={1},
  pages={L3},
  year={2019}
}

@article{event2019first3,
  title={First M87 event horizon telescope results. II. Array and instrumentation},
  author={Event Horizon Telescope Collaboration and Akiyama, Kazunori and Alberdi, Antxon and Alef, Walter and Asada, Keiichi and Azulay, Rebecca and Baczko, Anne-Kathrin and Ball, David and Balokovi{\'c}, Mislav and Barrett, John and others},
  journal={The Astrophysical Journal Letters},
  volume={875},
  number={1},
  pages={L2},
  year={2019},
  publisher={The American Astronomical Society}
}

@article{akiyama87event,
  title={Event Horizon Telescope Results. I},
  author={Akiyama, K},
  journal={The Shadow of the Supermassive Black Hole, First M},
  volume={87},
  pages={875},
  year={2019}
}

@article{akiyama2019first,
  title={First M87 event horizon telescope results. IV. Imaging the central supermassive black hole},
  author={Akiyama, Kazunori and Alberdi, Antxon and Alef, Walter and Asada, Keiichi and Azulay, Rebecca and Baczko, Anne-Kathrin and Ball, David and Balokovi{\'c}, Mislav and Barrett, John and Bintley, Dan and others},
  journal={The Astrophysical Journal Letters},
  volume={875},
  number={1},
  pages={L4},
  year={2019},
  publisher={IoP Publishing}
}

@article{event2019first,
  title={First M87 event horizon telescope results. V. Physical origin of the asymmetric ring},
  author={Event Horizon Telescope Collaboration and Akiyama, Kazunori and Alberdi, Antxon and Alef, Walter and Asada, Keiichi and Azulay, Rebecca and Baczko, Anne-Kathrin and Ball, David and Balokovi{\'c}, Mislav and Barrett, John and others},
  journal={The Astrophysical Journal Letters},
  volume={875},
  number={1},
  pages={L5},
  year={2019},
  publisher={The American Astronomical Society}
}

@article{event2019first2,
  title={First M87 event horizon telescope results. VI. The shadow and mass of the central black hole},
  author={Event Horizon Telescope Collaboration and Akiyama, Kazunori and Alberdi, Antxon and Alef, Walter and Asada, Keiichi and Azulay, Rebecca and Baczko, Anne-Kathrin and Ball, David and Balokovi{\'c}, Mislav and Barrett, John and others},
  journal={The Astrophysical Journal Letters},
  volume={875},
  number={1},
  pages={L6},
  year={2019},
  publisher={The American Astronomical Society}
}

@article{psaltis2019testing,
  title={Testing general relativity with the Event Horizon Telescope},
  author={Psaltis, Dimitrios},
  journal={General Relativity and Gravitation},
  volume={51},
  number={10},
  pages={137},
  year={2019},
  publisher={Springer}
}

@article{perlick2022calculating,
  title={Calculating black hole shadows: Review of analytical studies},
  author={Perlick, Volker and Tsupko, Oleg Yu},
  journal={Physics Reports},
  volume={947},
  pages={1--39},
  year={2022},
  publisher={Elsevier}
}

@article{amarilla2012shadow,
  title={Shadow of a rotating braneworld black hole},
  author={Amarilla, Leonardo and Eiroa, Ernesto F},
  journal={Physical Review D—Particles, Fields, Gravitation, and Cosmology},
  volume={85},
  number={6},
  pages={064019},
  year={2012},
  publisher={APS}
}

@article{eiroa2018shadow,
  title={Shadow cast by rotating braneworld black holes with a cosmological constant},
  author={Eiroa, Ernesto F and Sendra, Carlos M},
  journal={The European Physical Journal C},
  volume={78},
  number={2},
  pages={91},
  year={2018},
  publisher={Springer}
}

@article{papnoi2014shadow,
  title={Shadow of five-dimensional rotating Myers-Perry black hole},
  author={Papnoi, Uma and Atamurotov, Farruh and Ghosh, Sushant G and Ahmedov, Bobomurat},
  journal={Physical Review D},
  volume={90},
  number={2},
  pages={024073},
  year={2014},
  publisher={APS}
}

@article{singh2018shadow,
  title={Shadow of Schwarzschild--Tangherlini black holes},
  author={Singh, Balendra Pratap and Ghosh, Sushant G},
  journal={Annals of Physics},
  volume={395},
  pages={127--137},
  year={2018},
  publisher={Elsevier}
}

@article{amir2018shadows,
  title={Shadows of rotating five-dimensional charged EMCS black holes},
  author={Amir, Muhammed and Singh, Balendra Pratap and Ghosh, Sushant G},
  journal={The European Physical Journal C},
  volume={78},
  number={5},
  pages={399},
  year={2018},
  publisher={Springer}
}

@article{belhaj2020deflection,
  title={Deflection angle and shadow behaviors of quintessential black holes in arbitrary dimensions},
  author={Belhaj, A and Benali, M and El Balali, A and El Moumni, H and Ennadifi, SE},
  journal={Classical and Quantum Gravity},
  volume={37},
  number={21},
  pages={215004},
  year={2020},
  publisher={IOP Publishing}
}

@article{ferrari1984new,
  title={New approach to the quasinormal modes of a black hole},
  author={Ferrari, Valeria and Mashhoon, Bahram},
  journal={Physical Review D},
  volume={30},
  number={2},
  pages={295},
  year={1984},
  publisher={APS}
}

@article{leaver1985analytic,
  title={An analytic representation for the quasi-normal modes of Kerr black holes},
  author={Leaver, Edward W},
  journal={Proceedings of the Royal Society of London. A. Mathematical and Physical Sciences},
  volume={402},
  number={1823},
  pages={285--298},
  year={1985},
  publisher={The Royal Society London}
}

@article{cho2012new,
  title={A new approach to black hole quasinormal modes: a review of the asymptotic iteration method},
  author={Cho, HT and Cornell, AS and Doukas, Jason and Huang, T-R and Naylor, Wade},
  journal={Advances in Mathematical Physics},
  volume={2012},
  number={1},
  pages={281705},
  year={2012},
  publisher={Wiley Online Library}
}

@article{gundlach1994late,
  title={Late-time behavior of stellar collapse and explosions. I. Linearized perturbations},
  author={Gundlach, Carsten and Price, Richard H and Pullin, Jorge},
  journal={Physical Review D},
  volume={49},
  number={2},
  pages={883},
  year={1994},
  publisher={APS}
}

@article{berti2007mining,
  title={Mining information from binary black hole mergers: a comparison of estimation methods for complex exponentials in noise},
  author={Berti, Emanuele and Cardoso, Vitor and Gonzalez, Jose A and Sperhake, Ulrich},
  journal={Physical Review D—Particles, Fields, Gravitation, and Cosmology},
  volume={75},
  number={12},
  pages={124017},
  year={2007},
  publisher={APS}
}

@article{lin2017matrix,
  title={A matrix method for quasinormal modes: Schwarzschild black holes in asymptotically flat and (anti-) de Sitter spacetimes},
  author={Lin, Kai and Qian, Wei-Liang},
  journal={Classical and Quantum Gravity},
  volume={34},
  number={9},
  pages={095004},
  year={2017},
  publisher={IOP Publishing}
}

@article{schutz1985black,
  title={Black hole normal modes: a semianalytic approach},
  author={Schutz, Bernard F and Will, Clifford M},
  journal={The Astrophysical Journal},
  volume={291},
  pages={L33--L36},
  year={1985}
}

@article{kokkotas1988black,
  title={Black-hole normal modes: A WKB approach. III. The Reissner-Nordstr{\"o}m black hole},
  author={Kokkotas, Kostas D and Schutz, Bernard F},
  journal={Physical Review D},
  volume={37},
  number={12},
  pages={3378},
  year={1988},
  publisher={APS}
}

@article{iyer1987black,
  title={Black-hole normal modes: A WKB approach. I. Foundations and application of a higher-order WKB analysis of potential-barrier scattering},
  author={Iyer, Sai and Will, Clifford M},
  journal={Physical Review D},
  volume={35},
  number={12},
  pages={3621},
  year={1987},
  publisher={APS}
}

@article{konoplya2011quasinormal,
  title={Quasinormal modes of black holes: From astrophysics to string theory},
  author={Konoplya, Roman A and Zhidenko, Alexander},
  journal={Reviews of Modern Physics},
  volume={83},
  number={3},
  pages={793--836},
  year={2011},
  publisher={APS}
}

@article{ashtekar2006quantum,
  title={Quantum nature of the big bang},
  author={Ashtekar, Abhay and Pawlowski, Tomasz and Singh, Parampreet},
  journal={Physical review letters},
  volume={96},
  number={14},
  pages={141301},
  year={2006},
  publisher={APS}
}

@article{vazquez2003strong,
  title={Strong field gravitational lensing by a Kerr black hole},
  author={Vazquez, Samuel E and Esteban, Ernesto P},
  journal={arXiv preprint gr-qc/0308023},
  year={2003}
}

@article{raza2024influence,
  title={Influence of plasma on the optical appearance of spinning black hole in Kalb-Ramond gravity and its Existence around M87* and Sgr A},
  author={Raza, Muhammad Ali and Zubair, M and Maqsood, Eiman},
  journal={Journal of Cosmology and Astroparticle Physics},
  volume={2024},
  number={05},
  pages={047},
  year={2024},
  publisher={IOP Publishing}
}

@article{atamurotov2023quantum,
  title={Quantum effects on the black hole shadow and deflection angle in the presence of plasma},
  author={Atamurotov, Farruh and Jamil, Mubasher and Jusufi, Kimet},
  journal={Chinese Physics C},
  volume={47},
  number={3},
  pages={035106},
  year={2023},
  publisher={Chinese Physical Society and the Institute of High Energy Physics of the~…}
}

@article{nozari2025investigating,
  title={Investigating QED effects on the thin accretion disk properties around rotating Euler--Heisenberg black holes},
  author={Nozari, Kourosh and Saghafi, Sara and Aliyan, Fatemeh},
  journal={The European Physical Journal C},
  volume={85},
  number={7},
  pages={735},
  year={2025},
  publisher={Springer}
}

@article{nozari2025accretion,
  title={Accretion onto a charged black hole in consistent 4D Einstein-Gauss-Bonnet gravity},
  author={Nozari, Kourosh and Saghafi, Sara and Hassani, Mohammad},
  journal={Journal of High Energy Astrophysics},
  volume={45},
  pages={214--230},
  year={2025},
  publisher={Elsevier}
}

@article{nozari2023asymptotically,
  title={Asymptotically locally flat and AdS higher-dimensional black holes of Einstein--Horndeski--Maxwell gravity in the light of EHT observations: shadow behavior and deflection angle},
  author={Nozari, Kourosh and Saghafi, Sara},
  journal={The European Physical Journal C},
  volume={83},
  number={7},
  pages={588},
  year={2023},
  publisher={Springer}
}

@article{aktar2025shadows,
  title={Shadows and strong gravitational lensing around black hole-like compact object in quadratic gravity},
  author={Aktar, Somi and Molla, Niyaz Uddin and Rahaman, Farook and Mustafa, G},
  journal={Journal of High Energy Astrophysics},
  volume={47},
  pages={100385},
  year={2025},
  publisher={Elsevier}
}

@article{zhong2021qed,
  title={QED effects on Kerr black hole shadows immersed in uniform magnetic fields},
  author={Zhong, Zhen and Hu, Zezhou and Yan, Haopeng and Guo, Minyong and Chen, Bin},
  journal={Physical Review D},
  volume={104},
  number={10},
  pages={104028},
  year={2021},
  publisher={APS}
}

@article{jafarzade2025optical,
  title={Optical signatures of Einstein--Euler--Heisenberg AdS/dS black holes in the light of event horizon telescope},
  author={Jafarzade, Khadije and Bazyar, Zeynab and Saghafi, Sara and Nozari, Kourosh},
  journal={The European Physical Journal C},
  volume={85},
  number={8},
  pages={869},
  year={2025},
  publisher={Springer}
}

@article{nozari2026rotating,
  title={Rotating Black Holes with Primary Scalar Hair: Shadow Signatures in Beyond Horndeski Gravity},
  author={Nozari, Kourosh and Hajebrahimi, Milad and Saghafi, Sara and Mustafa, G and Saridakis, Emmanuel N},
  journal={arXiv preprint arXiv:2602.16237},
  year={2026}
}

@article{vagnozzi2023horizon,
  title={Horizon-scale tests of gravity theories and fundamental physics from the Event Horizon Telescope image of Sagittarius A∗},
  author={Vagnozzi, Sunny and Roy, Rittick and Tsai, Yu-Dai and Visinelli, Luca and Afrin, Misba and Allahyari, Alireza and Bambhaniya, Parth and Dey, Dipanjan and Ghosh, Sushant G and Joshi, Pankaj S and others},
  journal={Classical and Quantum Gravity},
  volume={40},
  number={16},
  pages={165007},
  year={2023},
  publisher={IOP Publishing}
}

@article{battista2026shadow,
  title={Shadow signatures and energy accumulation in Lorentzian-Euclidean black holes},
  author={Battista, Emmanuele and Capozziello, Salvatore and Chen, Che-Yu},
  journal={Physical Review D},
  volume={113},
  number={10},
  pages={104039},
  year={2026},
  publisher={APS}
}

@article{capozziello2025null,
  title={Null geodesics, causal structure, and matter accretion in Lorentzian-Euclidean black holes},
  author={Capozziello, Salvatore and Battista, Emmanuele and De Bianchi, Silvia},
  journal={Physical Review D},
  volume={112},
  number={4},
  pages={044009},
  year={2025},
  publisher={APS}
}

@article{de2026confronting,
  title={Confronting eikonal and post-Kerr methods with numerical evolution of scalar field perturbations in spacetimes beyond Kerr},
  author={De Simone, Ciro and V{\"o}lkel, Sebastian H and Kokkotas, Kostas D and De Falco, Vittorio and Capozziello, Salvatore},
  journal={Physical Review D},
  volume={113},
  number={10},
  pages={104004},
  year={2026},
  publisher={APS}
}

@article{Zhang2016,
    author = {Zhang, Xiangdong},
    title = {Higher dimensional loop quantum cosmology},
    journal = {Eur. Phys. J. C},
    volume = {76},
    number = {7},
    pages = {395},
    year = {2016},
    doi = {10.1140/epjc/s10052-016-4249-8},
    eprint = {1506.05597},
    archivePrefix = {arXiv},
    primaryClass = {gr-qc}
}

@article{Penrose1965,
  author  = {Penrose, Roger},
  title   = {Gravitational Collapse and Space-Time Singularities},
  journal = {Physical Review Letters},
  volume  = {14},
  pages   = {57--59},
  year    = {1965},
  doi     = {10.1103/PhysRevLett.14.57}
}

@article{HawkingPenrose1970,
  author  = {Hawking, Stephen W. and Penrose, Roger},
  title   = {The Singularities of Gravitational Collapse and Cosmology},
  journal = {Proceedings of the Royal Society of London. Series A,
             Mathematical and Physical Sciences},
  volume  = {314},
  number  = {1519},
  pages   = {529--548},
  year    = {1970},
  doi     = {10.1098/rspa.1970.0021}
}

@misc{Wald1997,
  author        = {Wald, Robert M.},
  title         = {Gravitational Collapse and Cosmic Censorship},
  year          = {1997},
  eprint        = {gr-qc/9710068},
  archivePrefix = {arXiv},
  primaryClass  = {gr-qc}
}

@book{hawking2023large,
  title={The large scale structure of space-time},
  author={Hawking, Stephen W and Ellis, George FR},
  year={2023},
  publisher={Cambridge university press}
}

@article{Tsukamoto2014,
  author  = {Tsukamoto, Naoki and Kitamura, Takao and Nakajima, Koki and Asada, Hideki},
  title   = {Gravitational lensing in Tangherlini spacetime in the weak gravitational field and the strong gravitational field},
  journal = {Physical Review D},
  volume  = {90},
  number  = {6},
  pages   = {064043},
  year    = {2014},
  doi     = {10.1103/PhysRevD.90.064043},
  eprint  = {1402.6823},
  archivePrefix = {arXiv},
  primaryClass  = {gr-qc}
}
\end{document}